\documentclass[10pt,journal,compsoc,onecolumn]{IEEEtran}
\usepackage{caption}
\usepackage[nocompress]{cite}
\usepackage{graphicx}
\usepackage{float}
\usepackage{amssymb,amsfonts,amsmath,amsthm}
\usepackage{breqn}
\usepackage{algorithm,algpseudocode}
\usepackage{enumitem}
\usepackage[hidelinks]{hyperref}
\usepackage{tabularx}
\usepackage{ragged2e}
\usepackage{longtable}
\usepackage{tikz}
\usepackage{ctable}

\usetikzlibrary{arrows,shapes,positioning,shadows,trees, calc}
\definecolor{myyellow}{HTML}{ecb348}
\definecolor{myorange}{HTML}{e45a35}
\definecolor{mygreen}{HTML}{8cb74c}
\definecolor{mygreenblue}{HTML}{5fb7a1}
\definecolor{myblue}{HTML}{3081ac}
\definecolor{mygray}{HTML}{a5a5a5}

\tikzset{
	basic/.style  = {draw, minimum height=0.7cm, text width=5cm, align=center, drop shadow, font=\sffamily, rectangle, rounded corners},
	root/.style   = {basic, draw=myblue, fill=myblue, text=white},
	level 2/.style = {basic, text width=5.5cm, draw=mygreenblue!40, fill=mygreenblue!40, text=myblue},
	level 3/.style = {basic, align=left, draw=myorange, fill=myorange, text=white}
}
\DeclareRobustCommand*{\IEEEauthorrefmark}[1]{%
	\raisebox{0pt}[0pt][0pt]{\textsuperscript{\footnotesize #1}}%
}

\begin{document}
	\bstctlcite{IEEEexample:BSTcontrol}
	\title{A Survey of Latent Factor Models in Recommender Systems}
	
	\author{Hind~I.~Alshbanat\IEEEauthorrefmark{1}, Hafida~Benhidour\IEEEauthorrefmark{1}, Said~Kerrache\IEEEauthorrefmark{1}
		\IEEEcompsocitemizethanks{\IEEEcompsocthanksitem \IEEEauthorrefmark{1} Computer Science Department, King Saud University, Riyadh 11543, Saudi Arabia.\\ \protect
			E-mail: hbenhidour@ksu.edu.sa}
	}
	
	\IEEEtitleabstractindextext{
		\begin{abstract}
			Recommender systems are essential tools in the digital era, providing personalized content to users in areas like e-commerce, entertainment, and social media. Among the many approaches developed to create these systems, latent factor models have proven particularly effective. This survey systematically reviews latent factor models in recommender systems, focusing on their core principles, methodologies, and recent advancements. The literature is examined through a structured framework covering learning data, model architecture, learning strategies, and optimization techniques. The analysis includes a taxonomy of contributions and detailed discussions on the types of learning data used, such as implicit feedback, trust, and content data, various models such as probabilistic, nonlinear, and neural models, and an exploration of diverse learning strategies like online learning, transfer learning, and active learning. Furthermore, the survey addresses the optimization strategies used to train latent factor models, improving their performance and scalability. By identifying trends, gaps, and potential research directions, this survey aims to provide valuable insights for researchers and practitioners looking to advance the field of recommender systems.
		\end{abstract}
		
		\begin{IEEEkeywords}
			Personalized Recommendations, Implicit Feedback, Trust Data, Nonlinear Models, Deep Neural Networks, Self-Supervised Learning, Transfer Learning, Optimization Techniques, Stochastic Gradient Descent, Data Sparsity, Scalability
		\end{IEEEkeywords}
	}
	
	\maketitle
	
	\IEEEdisplaynontitleabstractindextext
	
	\IEEEpeerreviewmaketitle
	
	\section{Introduction}
	\label{sec:introduction}
	Recommender systems are indispensable in the digital age. They play a crucial role in filtering large amounts of data to provide personalized content to users. These systems assist users in making informed choices across various domains, such as product selection, movie preferences, and music discovery. Recommender systems boost user satisfaction and engagement by forecasting user preferences using historical data and contextual information.
	
	Latent factor models have emerged as highly effective approaches for constructing recommender systems. These models harness the potential of matrix factorization techniques to capture the underlying patterns in user-item interactions. By representing users and items within a shared latent space, they address the sparsity and scalability challenges inherent in recommendation tasks.
	
	\begin{figure}[!ht]
		\centering
		\includegraphics[width=0.8\textwidth]{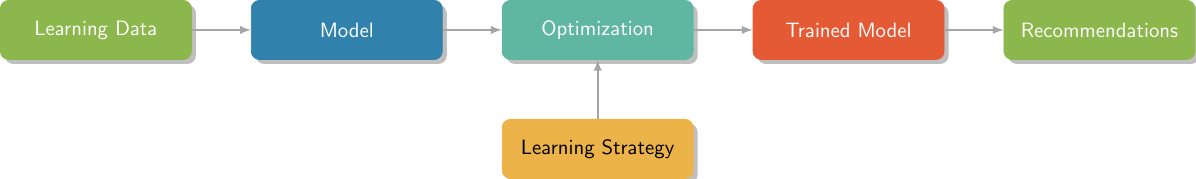}
		\caption{The structure of a latent factor model based recommender system.}
		\label{fig:rs}
	\end{figure}
	
	To better understand the inner workings of latent factor models in recommender systems, we present a general structure in Figure~\ref{fig:rs}. This structure highlights several key components that are essential for the effective operation of these systems:
	
	\begin{itemize}
		\item \textbf{Learning Data}: This includes the input data used for training the model. Learning data forms the foundation of the recommender system and typically consists of user-item interaction records. These records can include explicit feedback, such as ratings, or implicit feedback, such as clicks, views, and purchase history. The quality and quantity of this data significantly influence the performance of the recommender system.
		
		\item \textbf{Model}: The structure that encodes user-item interactions. The model captures the relationships between users and items, translating the complex patterns in the interaction data into a mathematical representation. This representation allows the system to predict user preferences for unseen items.
		
		\item \textbf{Learning Strategy}: The approach used to train the model. The learning strategy defines how the model is exposed to the data, how it learns from it, and how it iteratively improves its performance. It includes decisions on the type of learning (e.g., batch, online) and specific techniques that guide the model training process to ensure effective learning from the data.
		
		\item \textbf{Optimization}: The process of tuning the model parameters to minimize prediction error. Optimization involves adjusting the model's parameters to improve its accuracy in predicting user preferences. This process uses various algorithms to find the best parameter values that minimize the difference between the predicted and actual user interactions.
		
		\item \textbf{Trained Model}: The output of the optimization process, which can generate recommendations. Once the model parameters are optimized, the trained model can produce accurate and personalized recommendations based on the learned patterns in the data.
		
		\item \textbf{Recommendations}: The trained model calculates the final output provided to the user. Recommendations are the end product of the recommender system, presenting users with personalized content suggestions that are most likely to match their preferences and needs.
	\end{itemize}
	
	Given the complexity and rapid advancements in recommender system technologies, a comprehensive survey is necessary to gather and summarize current knowledge. While existing surveys have significantly contributed to our understanding of specific advancements, such as integrating deep learning with latent factor models \cite{TegeneAbebe2023DLaE} or exploring deeper versions of latent factor models using deep learning \cite{MongiaAanchal2020Dlfm}, our survey aims to provide a more holistic view of the role of latent factor models in recommender systems (see Table \ref{table:surveys}).
	
	\begin{table}[!t]
		\caption{Comparison to recent surveys on recommender systems.}
		\label{table:surveys}
		\centering
		\begin{tabular}{l c p{7cm} p{7cm}}
			\toprule
			\textbf{Reference} & \textbf{Year} & \textbf{Title} & \textbf{Difference with the current survey} \\ \midrule
			\cite{YangLi2023RDiR} & 2023 & Recent Developments in Recommender Systems: A Survey & Focuses on recent developments across various models, not specifically on latent factor models. \\ \midrule
			\cite{LiuQidong2023MRSA} & 2023 & Multimodal Recommender Systems: A Survey & Concentrates on multimodal data in recommender systems, whereas the current survey focuses on latent factor models. \\ \midrule
			\cite{CasilloMario2023Crsa} & 2023 & Context-aware recommender systems and cultural heritage: A survey & Focuses on context-aware systems and cultural heritage applications, different from the latent factor model focus. \\ \midrule
			\cite{wu2022graph} & 2022 & Graph Neural Networks in Recommender Systems: A Survey & Surveys graph neural networks in recommender systems, whereas the current survey is about latent factor models. \\ \midrule
			\cite{yu2023selfsupervised} & 2023 & Self-Supervised Learning for Recommender Systems: A Survey & Focuses on self-supervised learning techniques in recommender systems, unlike the current survey on latent factor models. \\ \bottomrule
		\end{tabular}
	\end{table}
	
	Specifically, we examine the literature through a structured framework that considers learning data, model architecture, learning strategies, and optimization techniques, offering a comprehensive and organized review of the field. Additionally, the survey addresses the need for a unified methodology for categorizing contributions in latent factor models across different applications and datasets. It offers a taxonomy that organizes contributions across multiple dimensions, highlighting individual papers' strengths and identifying trends and research gaps.
	
	Our methodology in this survey involves a systematic examination of the literature through the framework presented in Figure~\ref{fig:rs}. We evaluate related work from four perspectives: the learning data used, the model architecture, the learning strategy employed, and the optimization algorithm applied. This taxonomy of contributions is not mutually exclusive, and some papers may offer advancements in multiple categories. For readability, however, we present each paper in the category with the most substantial contribution.
	
	The rest of the paper is divided into two parts. Sections~\ref{sec:rs} and \ref{sec:lfm} cover background material on recommender systems and latent factor models. Section~\ref{sec:rs} covers the fundamentals of recommender systems, providing an overview of collaborative filtering, content-based filtering, and hybrid systems. It discusses the strengths and weaknesses of each approach and their application scenarios. Section~\ref{sec:lfm} introduces latent factor models, explaining their basic principles, introducing mathematical notation, and presenting the fundamental algorithms used to develop and optimize these models for recommender systems.
	
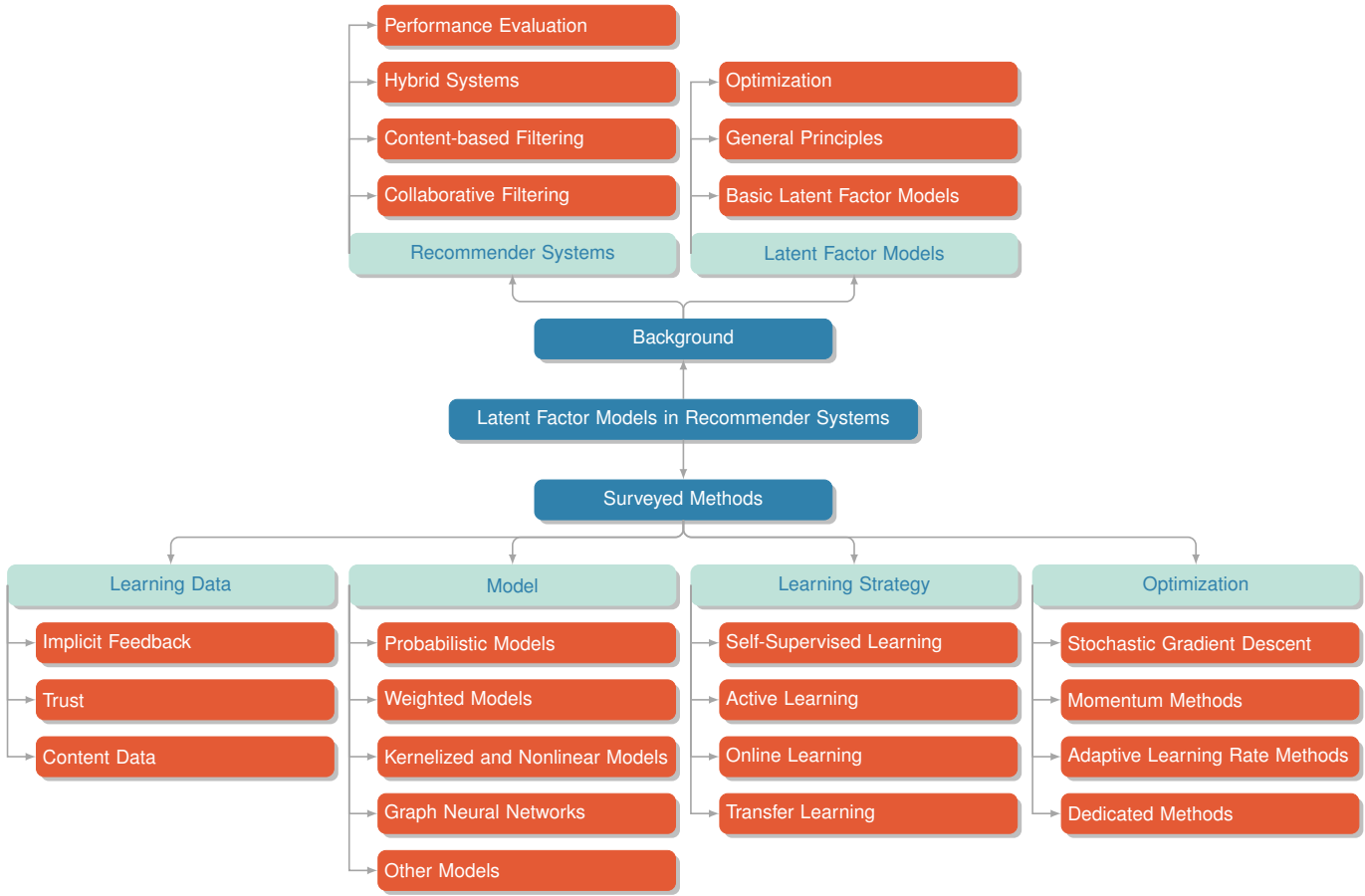
\begin{figure}[!t]
		\centering
		\resizebox{\textwidth}{!}{
			\begin{tikzpicture}[
				level 1/.style={sibling distance=6cm},
				edge from parent/.style={->,thick, draw=mygray},
				>=latex]
				
				\node[root, minimum width=8cm, text width=8cm, draw=myblue, fill=myblue] (paper) {\hyperref[sec:introduction]{Latent Factor Models in Recommender Systems}}
				child [edge from parent path={(\tikzparentnode.south) -- ++(0,-0.3cm) -| (\tikzchildnode)}]{node[root, below=0.7cm of paper, draw=myblue, fill=myblue] (survey) {\hyperref[sec:learning-data]{Surveyed Methods}}
					child {node[level 2] (c1) {\hyperref[sec:learning-data]{Learning Data}}}
					child {node[level 2] (c2) {\hyperref[sec:model]{Model}}}
					child {node[level 2] (c3) {\hyperref[sec:learning-strategy]{Learning Strategy}}}
					child {node[level 2] (c4) {\hyperref[sec:optimization]{Optimization}}}
				}
				child[grow=up, edge from parent path={(\tikzparentnode.north) -- ++(0,0.3cm) -| (\tikzchildnode)}]{node[root, above=0.7cm of paper, draw=myblue, fill=myblue] (fundamentals) {\hyperref[sec:rs]{Background}}
					child {node[level 2] (r2) {\hyperref[sec:lfm]{Latent Factor Models}}}
					child {node[level 2] (r1) {\hyperref[sec:rs]{Recommender Systems}}}
				}
				
				node[level 3, below=0.3cm of c1.south east, anchor=north east] (c11) {\hyperref[sec:implicit-feedback]{Implicit Feedback}}
				node[level 3, below of=c11] (c12) {\hyperref[sec:trust]{Trust}}
				node[level 3, below of=c12] (c13) {\hyperref[sec:content-data]{Content Data}}
				node[level 3, below=0.3cm of c2.south east, anchor=north east] (c21) {\hyperref[sec:probabilistic-models]{Probabilistic Models}}
				node[level 3, below of=c21] (c22) {\hyperref[sec:weighted-models]{Weighted Models}}
				node[level 3, below of=c22] (c23) {\hyperref[sec:kernelized-and-nonlinear-models]{Kernelized and Nonlinear Models}}
				node[level 3, below of=c23] (c24) {\hyperref[sec:gnn]{Graph Neural Networks}}
				node[level 3, below of=c24] (c25) {\hyperref[sec:other-models]{Other Models}}
				node[level 3, below=0.3cm of c3.south east, anchor=north east] (c31) {\hyperref[sec:ssl]{Self-Supervised Learning}}
				node[level 3, below of=c31] (c32) {\hyperref[sec:active-learning]{Active Learning}}
				node[level 3, below of=c32] (c33) {\hyperref[sec:online-learning]{Online Learning}}
				node[level 3, below of=c33] (c34) {\hyperref[sec:transfer-learning]{Transfer Learning}}
				node[level 3, below=0.3cm of c4.south east, anchor=north east] (c41) {\hyperref[sec:sgd]{Stochastic Gradient Descent}}
				node[level 3, below of=c41] (c42) {\hyperref[sec:momentum]{Momentum Methods}}
				node[level 3, below of=c42] (c43) {\hyperref[sec:alr]{Adaptive Learning Rate Methods}}
				node[level 3, below of=c43] (c44) {\hyperref[sec:dedicated]{Dedicated Methods}}
				node[level 3, above=0.3cm of r1.north east, anchor=south east] (r11) {\hyperref[sec:cf]{Collaborative Filtering}}
				node[level 3, above of=r11] (r12) {\hyperref[sec:cbf]{Content-based Filtering}}
				node[level 3, above of=r12] (r13) {\hyperref[sec:hs]{Hybrid Systems}}
				node[level 3, above of=r13] (r14) {\hyperref[sec:performance]{Performance Evaluation}}
				node[level 3, above=0.3cm of r2.north east, anchor=south east] (r21) {\hyperref[sec:blfm]{Basic Latent Factor Models}}
				node[level 3, above of=r21] (r22) {\hyperref[sec:gp]{General Principles}}
				node[level 3, above of=r22] (r23) {\hyperref[sec:opt]{Optimization}}
				; 
				
				\draw[->,thick, draw=mygray] (c1.180) |- (c11);
				\draw[->,thick, draw=mygray] (c1.180) |- (c12);
				\draw[->,thick, draw=mygray] (c1.180) |- (c13);
				\draw[->,thick, draw=mygray] (c2.180) |- (c21);
				\draw[->,thick, draw=mygray] (c2.180) |- (c22);
				\draw[->,thick, draw=mygray] (c2.180) |- (c23);
				\draw[->,thick, draw=mygray] (c2.180) |- (c24);
				\draw[->,thick, draw=mygray] (c2.180) |- (c25);
				
				\draw[->,thick, draw=mygray] (c3.180) |- (c31);
				\draw[->,thick, draw=mygray] (c3.180) |- (c32);
				\draw[->,thick, draw=mygray] (c3.180) |- (c33);
				\draw[->,thick, draw=mygray] (c3.180) |- (c34);
				
				\draw[->,thick, draw=mygray] (c4.180) |- (c41);
				\draw[->,thick, draw=mygray] (c4.180) |- (c42);
				\draw[->,thick, draw=mygray] (c4.180) |- (c43);
				\draw[->,thick, draw=mygray] (c4.180) |- (c44);
				
				\draw[->,thick, draw=mygray] (r1.180) |- (r11);
				\draw[->,thick, draw=mygray] (r1.180) |- (r12);
				\draw[->,thick, draw=mygray] (r1.180) |- (r13);
				\draw[->,thick, draw=mygray] (r1.180) |- (r14);
				
				\draw[->,thick, draw=mygray] (r2.180) |- (r21);
				\draw[->,thick, draw=mygray] (r2.180) |- (r22);
				\draw[->,thick, draw=mygray] (r2.180) |- (r23);				
			\end{tikzpicture}
		}
		\caption{Overview of the structure of this survey.}
		\label{fig:struct}
	\end{figure}

	The second part of the paper, from Sections \ref{sec:learning-data} to \ref{sec:optimization}, contains the surveyed methods. An overview of the structure of this document is shown in Figure~\ref{fig:struct}. Section~\ref{sec:learning-data} discusses the various types of learning data used to improve the quality of recommendations, including implicit feedback, trust, and content data. Section~\ref{sec:model} discusses various models used in latent factor approaches, including probabilistic, weighted, kernelized, and nonlinear models. It examines how these models enhance the capability and accuracy of recommender systems. Section~\ref{sec:learning-strategy} reviews learning strategies like online, transfer, and active learning, highlighting their applications and benefits in improving the adaptability and effectiveness of recommender systems. Section~\ref{sec:optimization} focuses on optimization strategies for latent factor models. It covers general-purpose algorithms such as stochastic gradient descent and its various variants and more specialized algorithms dedicated to latent factor models in recommender systems. 
	
	Each section of the survey concludes with an in-depth discussion of existing contributions and highlights interesting future research directions. By examining the state-of-the-art methods and identifying gaps in the current research, we aim to provide insights into potential areas for future work and guide researchers in advancing the field of recommender systems.
	
	\section{Fundamentals of Recommender Systems}
	\label{sec:rs}
	Recommender systems are software systems that automatically filter large amounts of data to find relevant information for the users \cite{taghipour2008hybrid}. They aim to support users in various decision-making processes, such as which products to buy or services to engage in. 
	Despite the large amount of online information, there are typically few rated items due to users being unwilling to provide ratings. As a result, recommendation data are typically highly sparse. The recommender systems' task is to predict unrated items based on a small portion of rated items. The system takes as input a user model (such as ratings, preferences, situational context, demographics) and items with or without a description and finds the corresponding relevance scores \cite{jannach2010recommender}. Finally, it returns a list of recommended items relevant to the target user.
	
	Several approaches have been proposed to solve the recommendation problem. These approaches are mainly organized into three classes \cite{adomavicius2005toward, jannach2010recommender}: 
	\begin{itemize}
		\item Collaborative filtering recommender systems are based on user profiles and ratings of similar users and recommend items to target users that other users with similar tastes have previously preferred.
		
		\item Content-based recommender systems exploit user profiles and item descriptions to recommend items to the target users that are similar to what they liked in the past.
		
		\item Hybrid recommender systems combine collaborative filtering and content-based approaches to improve recommendation accuracy and diversity.
	\end{itemize}
	
	The rest of this section will present each category, detailing the most common and basic algorithms associated with each and their respective advantages and disadvantages. Specifically, we will delve into collaborative filtering, starting with its memory-based and model-based variants, and elaborate on the fundamental concepts, methodologies, and critical differences between user-based and item-based approaches. Following that, we will explore content-based recommender systems, highlighting how these systems utilize item features and user preferences to generate personalized recommendations. Finally, we will examine hybrid recommender systems, which combine the strengths of collaborative filtering and content-based methods to mitigate their limitations and enhance recommendation quality.
	
	\subsection{Collaborative Filtering}
	\label{sec:cf}
	Collaborative filtering recommender systems are the most familiar and widely used personalized recommendation algorithms. The idea of collaborative filtering is based on the premise that users with similar tastes in the past tend to like similar items in the future \cite{schafer2007collaborative}. These systems first detect user similarities based on historical data and then use these similarities to recommend new items. Collaborative filtering is primarily based on ratings, whether explicit (e.g., numeric, thumbs up or down) or implicit (e.g., clicking, watching). 
	Collaborative filtering is divided into two categories: memory-based and model-based collaborative filtering \cite{breese2013empirical}. Memory-based collaborative filtering requires a complete user–item database in memory for computing recommendations. In contrast, model-based collaborative filtering requires only an abstract representation of such a database.
	
	\subsubsection{Memory-based Collaborative Filtering}
	Memory-based collaborative filtering uses similarity measures and various prediction computation techniques to estimate unknown ratings. In general, memory-based collaborative filtering algorithms work as follows:
	\begin{enumerate}
		\item Build a user-item rating matrix where rows correspond to users and columns correspond to items. An entry $r_{ui}$ in this matrix contains the rating assigned to item $i$ by user $u$.
		\item Calculate the similarity row-wise between users or column-wise between items using a prescribed similarity measure.
		\item Select the $k$ users or items most similar to the target user or item.
		\item Predict unknown ratings using a specified prediction calculation method.
	\end{enumerate}
	
	Memory-based collaborative filtering is divided into user-based and item-based collaborative filtering \cite{singh2019comparative, santos2014extending} depending on whether the similarity computation in Step 2 is conducted on users or items. The idea of user-based collaborative filtering algorithms is that users who have rated the same items similarly in the past have similar interests. Thus, the system can infer the future rating the target user will give to a given item based on the rating given to it by similar users. Several similarity measures can be used to compute the similarity between the users, including cosine similarity \cite{adomavicius2011context, billsus1998learning, billsus2000user, lang1995newsweeder}, Pearson's Correlation Coefficient (PCC) \cite{koohi2017new, resnick1994grouplens}, Spearman's correlation coefficient \cite{herlocker1999algorithmic}, Adjusted cosine similarity \cite{sarwar2001item}, and the Jaccard coefficient \cite{liu2014new}. Within this plethora of measures, cosine similarity and the Pearson correlation coefficient are the most widely used in practice. Cosine similarity considers the user's ratings as an $n$-dimensional vector and computes user similarity through the angles between vectors:
	\begin{equation}
		\label{eq:cosine similarity}
		sim_{uv}=cos(r_u,r_v)=\frac{r_u \cdot r_v}{\left\|r_u\right\|_2 \left\|r_v \right\|_2} = \frac{\sum_{i\in I_{uv}}r_{ui} r_{vi}}{\sqrt{\sum_{i\in I_u}r^2_{ui}}\sqrt{\sum_{i\in I_v}r^2_{vi}}},
	\end{equation}
	where $sim_{uv}$ represents the similarity between user $u$ and user $v$, $r_u$ and $r_v$ represent the rating vectors of users $u$ and $v$, ${\left\| \cdot \right\|_2}$ denotes the $L_2$ norm, $r_{ui}$ and $r_{vi}$ represent ratings of users $u$ and $v$ on item $i$, $I_u$ and $I_v$ are the set of items already rated by user $u$ and user $v$, respectively, and $I_{uv}$ stands for the shared items rated by both users.
	
	Pearson's correlation coefficient, on the other hand, measures the similarity between users based on the linear correlation between their shared ratings:
	\begin{equation}
		\label{eq:PCC}
		sim_{uv}=\frac{\sum_{i\in I_{uv}}(r_{ui}-\bar{r}_u)(r_{vi}-\bar{r}_v)}{{\sqrt{\sum_{i\in I_{uv}}(r_{ui}-\bar{r}_u)}}{\sqrt{\sum_{i\in I_{uv}}(r_{vi}-\bar{r}_v)}}},
	\end{equation}
	where $\bar{r}_u$ and ${\bar{r}_v}$  denote the average ratings user $u$ and $v$ respectively.
	
	Once the $k$ nearest neighbor or the highest similar users to the target user have been selected, user-based collaborative filtering then predicts unknown ratings using the following formula \cite{moradi2015reliability}:
	\begin{equation}
		\label{eq:prediction}
		\hat{r}_{ui}=\bar{r}_u+\frac{\sum_{v\in N_u}sim_{uv}(r_{vi}-\bar{r}_v)}{\sum_{v\in N_u}\left|sim_{uv} \right|},
	\end{equation}
	where $N_u$ is the set of nearest neighbors of user $u$. Equation \eqref{eq:prediction} states that the predicted difference between the rating given to item $i$ and the average rating of user $u$ is the weighted sum of the deviations of the ratings given to the same item by the $k$  nearest neighbors of $u$. The absolute value is necessary here since some similarity measures may take a negative sign.
	
	Item-based collaborative filtering is similar to user-based collaborative filtering, except it computes the similarity between items rather than users. For example, Equation \eqref{eq:itemPCC} shows how to calculate the similarity between items using the Pearson correlation coefficient \cite{salter2006cinemascreen}:
	\begin{equation}
		\label{eq:itemPCC}
		sim_{ij}= \frac{\sum_{u\in U_{ij}}(r_{ui}-\bar{r}_u)(r_{uj}-\bar{r}_j)}{\sqrt{\sum_{u\in U_{ij}}(r_{ui}-\bar{r}_u)^2}\sqrt{\sum_{u\in U_{ij}}(r_{uj}-\bar{r}_j)^2}},
	\end{equation}
	where $sim_{ij}$ is the similarity of item $i$ and item $j$, $U_i$ and $U_j$ represent the set of users who rated item $i$ and item $j$, respectively, $U_{ij}$ is the set of users who rated both items, and $\bar{r}_u$ and $\bar{r}_j$ are the average ratings of item $i$ and item $j$, respectively.
	
	\subsubsection{Model-based Collaborative Filtering}
	Memory-based recommender systems are easy to understand and implement but unsuitable for large datasets due to their large storage requirements and poor computational efficiency and scalability. Model-based collaborative filtering avoids this drawback by building and learning a model based on the user-item rating matrix and then predicting unknown ratings \cite{ju2013new, salah2016dynamic}. Model-based collaborative filtering approaches can be classified into latent factor models \cite{koren2009matrix,ranjbar2015imputation, huang2016collaborative, bobadilla2017recommender, yuan2019singular}, clustering models \cite{o1999clustering}, Bayesian classifiers \cite{park2007location}, and various probabilistic relational models \cite{getoor1999using}. In this survey, we will focus on latent factor models because of their remarkable success in solving the recommendation problem, particularly their ability to handle sparsity and capture underlying patterns in the rating data. We will discuss these models in more detail in the remainder of this text.
	
	\subsection{Content-based Filtering}
	\label{sec:cbf}
	Content-based recommender systems provide recommendations to target users by comparing the representation of the description of items with the representation of the content of interest to the target user \cite{van2000using, lops2011content}. For example, if the user prefers action movies, the system learns to recommend other movies in the action genre.
	Content-based recommender systems can also extract information from the user's profile, such as gender, age, nationality, and demographic information, to improve recommendations \cite{pazzani1997learning, herlocker2004evaluating}. This information can prove valuable to the recommendation algorithm as it helps build a model of user preferences for their profile data.
	
	A content-based system analyzes and extracts meaningful features from item descriptions or user profiles using text mining and natural language processing techniques to generate recommendations. Using these features, a model is built or trained to capture the similarity between items or between items and the user's preferences. The system ranks the items based on their relevance, calculated using the learned model, and suggests the top recommendations to the user.
	
	Content-based recommender systems can effectively recommend items that users will like, even if those items are new or less popular. However, they tend to recommend items similar to those the user has already interacted with, which can result in a lack of serendipity in the recommendations. Their performance also depends heavily on the availability of informative, accurate item descriptions. However, these descriptions are often prone to human error, imperfections, or missing information, which can profoundly impact the quality of the recommendations.
	
	\subsection{Hybrid Systems}
	\label{sec:hs}
	To overcome the shortcomings of a single recommendation model, collaborative filtering and content-based algorithms can be combined in hybrid systems to improve the prediction accuracy of the recommender system \cite{melville2002content, chen2010hybrid, huang2016collaborative, miranda1999combining}. Combining collaborative and content-based algorithms into a hybrid recommender system can be achieved in various ways. One approach involves implementing each model separately and then aggregating their predictions to leverage the strengths of the base models. Another strategy incorporates characteristics of content-based recommendations into collaborative filtering or vice versa, resulting in a system that benefits from the unique advantages of both methods. A more integrated approach also involves building a fused model that inherently includes collaborative filtering and content-based characteristics.
	
	\subsection{Performance Evaluation}
	\label{sec:performance}
	The type of performance measures used to evaluate a recommender system depends on the query the system is designed to answer. The most common use case requires the system to predict a numerical rating given by a user to a specific item. The task is then considered a regression task, and usual regression performance measures are used. These typically include the Mean Absolute Error (MAE) \cite{breese2013empirical} and the Root Mean Square Error (RMSE) \cite{herlocker2004evaluating}. MAE computes the average absolute difference between predicted and actual ratings:
	\begin{equation}
		\label{eq:MAE}
		MAE = \frac{\sum_{r_{ui}\in \mathcal{T}}\left| r_{ui} - \hat{r}_{ui} \right|}{\left| \mathcal{T} \right|},
	\end{equation}
	where $\mathcal{T}$ is the test set, $\left| \mathcal{T} \right|$ represents the total number of ratings in the test set, $r_{ui}$ and $\hat{r}_{ui}$ represent the actual and predicted ratings for user $u$ and item $i$, respectively. RMSE computes the square root of the average of the squared differences between predicted and actual ratings:
	\begin{equation}
		\label{eq:RMSE}
		RMSE = \sqrt{\frac{\sum_{r_{ui} \in \mathcal{T}}(r_{ui} - \hat{r}_{ui})^2}{\left| \mathcal{T} \right|}}.
	\end{equation}
	
	Lower values of MAE and RMSE indicate better performance by the model. These two measures are generally positively correlated, but MAE has the property of treating all error magnitudes as equal. In contrast, RMSE, on the other hand, penalizes large errors more than small ones. For example, with MAE, an error of 2 in a single example is equivalent to two errors of 1 in two different examples, whereas with RMSE, an error of 2 in a single example is considered more severe than two errors of 1 in two different examples.
	
	A recommender system can also be used to discriminate between items relevant to a given user and those irrelevant rather than producing a numerical rating. This scenario is a classification task that can be evaluated using binary classification performance measures such as accuracy, recall, precision, and the F1 measure.
	
	In various practical scenarios, such as online shopping and reservation websites, recommender systems sort the available items in descending order of relevance and present them to the user. Having the most relevant items at the top of the list is desirable. Ranking metrics are used to evaluate the performance of this type of system \cite{hijikata2014offline}, which include Mean Average Precision (MAP), Mean Reciprocal Rank (MRR), Discounted Cumulative Gain (DCG), Normalized Discounted Cumulative Gain (NDCG), and Hit Ratio (HR). MAP is commonly used to evaluate such recommender systems. MAP is found by taking the mean of all users' Average Precision (AP). The AP for a single user is calculated as the average precision at $k$ for all $k$ corresponding to relevant items within the recommendation list. This process captures both the precision of the recommendations and their ranking quality. MRR focuses on the rank of the most relevant item. The higher its rank, the closer the MRR is to its maximum of 1. DCG measures the overall quality of the recommendation list by giving a higher weight to relevant items placed at the top. In contrast, NDCG normalizes DCG against the ideal ranking order, providing a scale from 0 to 1, where 1 represents the perfect order. HR is usually reported at a specific cut-off point, HR@$ k $, indicating if the relevant item is within the top $ k $ recommendations.
	
	\section{Fundamentals of Latent Factor Models}
	\label{sec:lfm}
	Latent factor models are among the most successful model-based techniques for recommender systems \cite{fu2018novel}. These models often focus on factorizing the user-item rating matrix into two lower-dimensional latent spaces. The latent factors are derived from the observed ratings using well-established linear algebra or optimization methods. In some models, a reconstruction function, such as the inner product of the user and item latent vectors, is then used to predict unknown ratings. For other models, such as those based on neural networks, the reconstruction function is learned from data.
	This section covers the fundamentals of latent factor models. We begin by discussing the precursors to most modern models and then provide a general formulation of them. We conclude by highlighting the key optimization techniques used to fit these models to data.
	
	\subsection{Basic Latent Factor Models}
	\label{sec:blfm}
\begin{figure}[!t]
		\centering
		\includegraphics[width=\textwidth]{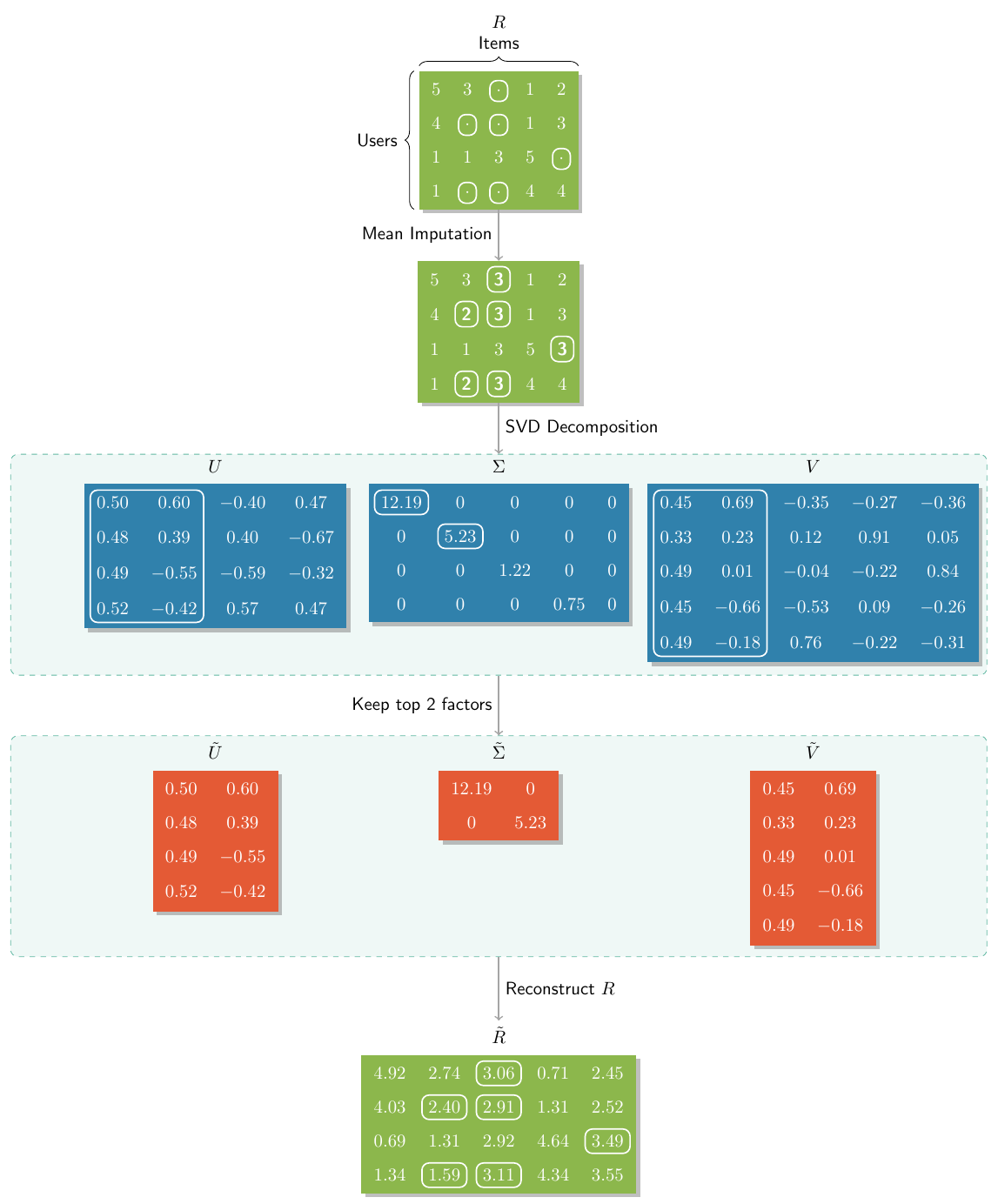}
		\caption{An example of using SVD to predict ratings. Initially, the rating matrix $R$ undergoes imputation, where missing ratings are replaced with the average rating of each item. Subsequently, $R$ is decomposed using SVD. The two factors corresponding to the largest singular values are retained to reconstruct an approximation $\tilde{R}$ of $R$.}
		\label{fig:svd-example}
	\end{figure}

	Singular Value Decomposition (SVD) \cite{sarwar2000application, zhou2015svd} is one of the earliest latent factor approaches for collaborative filtering. SVD is a classical linear algebra technique that generalizes eigenvalue decomposition to non-square matrices. More precisely, any $m \times n$ matrix $R$ can be decomposed as:
	\begin{equation}
		R = U \Sigma V^T,
	\end{equation}
	where $U$ is an $m \times m$ orthogonal matrix, $\Sigma$ is an $m \times n$ rectangular diagonal matrix (with all non-diagonal entries set to zero), and $V^T$ is the transpose of an $n \times n$ orthogonal matrix. The diagonal entries $\sigma_i = \Sigma_{ii}$ are known as the \emph{singular values} of $R$ and are guaranteed to be non-negative for a real matrix $R$. In collaborative filtering, $R$ represents the rating matrix, $U$ encapsulates user latent vectors, and $V$ contains item latent vectors. The magnitude of the singular values indicates each factor's contribution strength in explaining the observed ratings. SVD requires fully known matrices and, therefore, cannot be directly applied to rating matrices, as they typically contain missing values. This issue can be addressed through matrix imputation by filling in missing values with the mean rating of either the user or item (mean imputation) or setting them to zero (zero imputation). Nonetheless, applying SVD directly to an imputed matrix is not very useful, as the predicted rating will exactly equal the imputed values, causing the model to fail in generalization. Instead, we adopt the \emph{low-rank assumption}, which posits that the full high-dimensional matrix $R$ can be approximated by using a small number of factors that capture the data's essential structure and main patterns. More specifically, we retain the factors corresponding to the $k$ largest singular values and discard the rest, leading to the following approximation:
	\begin{equation}
		R \approx \tilde{R} = \tilde{U} \tilde{\Sigma} \tilde{V}^T,
	\end{equation}
	where $\tilde{U}$ is $m \times k$, $\tilde{\Sigma}$ is $k \times k$, and $\tilde{V}^T$ is $k \times n$. Figure~\ref{fig:svd-example} illustrates the SVD procedure for rating prediction.
	
	This approach to dimensionality reduction helps mitigate overfitting, enabling SVD to predict unseen ratings using a small number of factors. However, SVD faces several challenges. Primarily, rating matrices are typically sparse, and imputation introduces biases. For example, mean imputation can significantly reduce the ratings' variance, while zero imputation shifts the average rating towards lower values. Excessive imputation in highly sparse matrices also obscures existing patterns, resulting in poor generalization. Additionally, SVD incurs a high computational overhead from factoring the often high-dimensional rating matrix. These issues are addressed by modern latent factor models using various regularization and optimization techniques.
	
	In response to SVD's limitations, Matrix Factorization (MF) \cite{koren2009matrix} is a refined approach, optimizing the decomposition process for sparsity and computational efficiency.MF is a latent factor model that consists of factorizing the original user-item rating matrix $ R $ into two lower-dimensional matrices $ P $ and $ Q $:
	\begin{equation} 
		\label{eq:MF} 
		R \approx \tilde{R} = PQ^T, 
	\end{equation} 
	where $ P $ is an $ n \times k $ matrix representing the latent factors of $ n $ users, $ Q $ is an $ m \times k $ matrix representing the latent factors of $ m $ items, and $ k $ is the dimensionality of the latent factor space. Each row vector $ p_u $ of the matrix $ P $ constitutes the latent factor representation of user $ u $, whereas each row vector $ q_i $ of $ Q $ represents item $ i $. The unknown ratings are predicted by computing the dot product of the user and item latent vectors: 
	\begin{equation} 
		\label{eq:mf-pred} 
		\hat{r}_{ui}=\sum_{j=1}^{k}P_{u j} Q_{j i} = p_u^T q_i,
	\end{equation} 
	where $\hat{r}_{ui}$ refers to the predicted rating given by user $u$ to item $i$. Note that when considered individually, $ p_u $ and $ q_i $ are treated as column vectors. To find $p_u$ and $q_i$, the model is trained using a gradient descent algorithm, which minimizes the sum of squared errors (SSE) iteratively between predicted and observed ratings using the following function:
	\begin{equation} 
		\label{eq:mf-sse} 
		SSE = \dfrac{1}{2} \sum_{r_{ui} \in \mathcal{R}} e_{ui}^2= \dfrac{1}{2} \sum_{r_{ui} \in \mathcal{R}}\left (\hat{r}_{ui}-r_{ui}\right )^2 = \dfrac{1}{2} \sum_{r_{ui} \in \mathcal{R}}\left(p_u^T q_i-r_{ui}\right)^2,
	\end{equation} 
	where $\mathcal{R}$ is the set of known ratings, and $e_{ui}$ is the error between predicted and known ratings. At each iteration, the user and item vectors $p_u$ and $q_i$ are updated as follows: 
	\begin{align} 
		\label{eq:user} 
		p_u &= p_u + \eta \sum_{r_{ui} \in \mathcal{R}_u} e_{ui}q_i,\\
		q_i &= q_i + \eta  \sum_{r_{ui} \in \mathcal{R}_i} e_{ui}p_u,
	\end{align} 
	where $\mathcal{R}_u$ is the set of all ratings made by user $u$, $ \mathcal{R}_i $ is the set of all ratings of item $i$, and $\eta$, the learning rate, is an algorithm parameter determined empirically.
	
	For sparse data, training the model using the loss defined in Equation~\eqref{eq:mf-sse} may result in overfitting. Hence, the need to use $L_2$-norm regularization to reduce the model complexity by penalizing the norm of its parameters \cite{paterek2007improving}. The regularized objective function takes the form: 
	\begin{equation} 
		J_{MF} = \dfrac{1}{2} \sum_{r_{ui} \in \mathcal{R}}\left (p_u^T q_i - r_{ui}\right )^2 + \lambda \left(\sum_{u=1}^n \|p_u\|_2^2+ \sum_{i=1}^m \|q_i\|_2^2\right),
	\end{equation} 
	where $\lambda$ is the regularization coefficient, a hyper-parameter typically set by model selection techniques, $ \|p_u\|_2^2 = \sum_{j=1}^{k} p_{uj}^2$, and $ \|q_i\|_2^2 = \sum_{j=1}^{k} q_{ij}^2$.
	
	\subsection{General Principles}
	\label{sec:gp}
	Matrix Factorization constitutes a foundational framework applicable to many latent factor models. Generally, a latent factor model assumes that a set of latent factors can represent user preferences and item attributes. These factors enable the explanation of existing ratings and the prediction of future ones. Specifically, a latent factor model in the context of a recommender system with $n$ users and $m$ items can be defined as a tuple $\left(P, Q, w, f \right)$, where:
	\begin{itemize}
		\item $P=\{p_1, \ldots, p_n\}$ where $p_u \in \mathbb{R}^{k_u}$ represents the latent vector associated with user $i$, and $k_u$ is the dimension of the latent user space.
		
		\item $Q=\{q_1, \ldots, q_m\}$ where $q_i \in \mathbb{R}^{k_i}$ represents the latent vector associated with item $i$, and $k_i$ is the dimension of the latent item space.
		
		\item $w \in \mathbb{R}^d$ represents the model parameters shared among all users and items.
		
		\item $f$ is the reconstruction function used to predict ratings:
		\begin{equation}
			\hat{r}_{ui} = f(p_u, q_i, w).
		\end{equation}
		The reconstruction function $f$ can be as simple as a dot product, as in MF, or a complex nonlinear function represented by a deep neural network.
	\end{itemize} 
	For instance, in MF, the dimensions of the latent user space and the latent item space are equal ($k_u = k_i = k$); the model has no shared parameters, and the reconstruction function is given by Equation~\eqref{eq:mf-pred}.
	
	The model parameters, $P$, $Q$, and $w$, are determined by minimizing a loss function $\ell(\hat{r}_{ui}, r_{ui})$ that captures the error between the predicted rating $\hat{r}_{ui}$ and the actual rating $r_{ui}$. A popular choice for $\ell$ with numerical ratings is the squared error due to its nice analytical and numerical properties:
	\begin{equation}
		\ell(\hat{r}_{ui}, r_{ui}) = \frac{1}{2} \left(\hat{r}_{ui}-r_{ui}\right)^2.
	\end{equation}
	For binary ratings, binary cross-entropy is customarily used:
	\begin{equation}
		\ell(\hat{r}_{ui}, r_{ui}) = -r_{ui} \log(\hat{r}_{ui})-(1-r_{ui}) \log(1-\hat{r}_{ui}).
	\end{equation}
	The total loss over the training set is obtained by summing the losses of all ratings:
	\begin{equation}
		L(P, Q, w)= \sum_{r_{ui} \in \mathcal{R}} \ell(\hat{r}_{ui}, r_{ui}),
	\end{equation}  
	where $ \mathcal{R} $ is the set of all available ratings. Minimizing $L$ alone often leads to overfitting, especially in highly sparse datasets. Introducing a regularization term $\Omega$ biases the model towards simplicity, enhancing generalization. The regularized objective function is thus:
	\begin{equation}
		J(P, Q, w) = L(P, Q, w) + \lambda \Omega(P, Q, w),
	\end{equation} 
	where $\lambda$ is the regularization coefficient. Common regularization terms $\Omega$ include:
	\begin{itemize}
		\item $L_1$ Regularization (Lasso Regularization): penalizes the loss function with the $L_1$ norm of the model parameter vectors, encouraging sparsity:
		\begin{equation}
			\Omega(P, Q, w) = \sum_{u=1}^{n} \|p_u\|_1 + \sum_{i=1}^{m} \|q_i\|_1 + \|w\|_1.
		\end{equation}
		
		\item $L_2$ Regularization (Ridge Regularization): penalizes the loss function with the square of the $L_2$ norm of the model parameter vectors, leading to smoother solutions:
		\begin{equation}
			\Omega(P, Q, w) = \sum_{u=1}^{n} \|p_u\|_2^2 + \sum_{i=1}^{m} \|q_i\|_2^2 + \|w\|_2^2.
		\end{equation}
		This form of regularization is used in the original MF algorithm and is by far the most widely used due to its differentiability and convexity.
		
		\item Elastic Net Regularization: combines both $L_1$ and $L_2$ regularization to benefit from the properties of both, encouraging sparsity while also ensuring smoother solutions:
		\begin{equation}
			\Omega(P, Q, w) = \sum_{u=1}^{n} \|p_u\|_1 + \sum_{i=1}^{m} \|q_i\|_1 + \|w\|_1  + \sum_{u=1}^{n} \|p_u\|_2^2 + \sum_{i=1}^{m} \|q_i\|_2^2 + \|w\|_2^2.
		\end{equation}
	\end{itemize}
	In addition, methods like early stopping and dropout \cite{goodfellow} also help combat overfitting. Early stopping halts training once validation set performance worsens. Dropout, used in neural networks, randomly zeroes a fraction of outputs in a layer during training, introducing noise that helps the model generalize better.
	
	\subsection{Optimization}
	\label{sec:opt}
	Most latent factor approaches rely on continuous optimization algorithms to fit the model to training data. The fundamental algorithm used to find the optimum of a continuously differentiable function is gradient descent. To simplify notation, let us define the vector $ \theta $ by concatenating all model parameters:
	\begin{equation}
		\theta = \mathrm{concat}\left(p_1, \ldots, p_n, q_1, \ldots q_m, w\right).
	\end{equation}
	We combine all parameters into one vector because the optimization algorithms do not distinguish between the various model parameters. Gradient descent iteratively updates the parameter values, starting from an initial guess, using the gradient of the objective function:
	\begin{equation}
		\theta = \theta - \eta \nabla_{\theta} J(\theta),
	\end{equation}
	where $ \eta $ is a parameter known as the \emph{step size} or \emph{learning rate}. Note that the gradient is calculated with respect to the parameters $ \theta $, not the rating data $ r_{ui} $. A good choice of $ \eta $ is essential for the algorithm's convergence. Small values can result in slow convergence, while large values can cause oscillations or even divergence. The value of $ \eta $ is often considered a hyperparameter tuned using a validation set, although methods for automatically adjusting the learning rate during training also exist. It is important to remember that gradient descent is a local optimization algorithm, so the solution it finds may not be the global optimum. Good initialization strategies can help find better solutions. For example, restarting optimization from multiple starting points can mitigate local convergence issues.
	
\begin{figure}[!t]
		\centering
		\includegraphics[width=\textwidth]{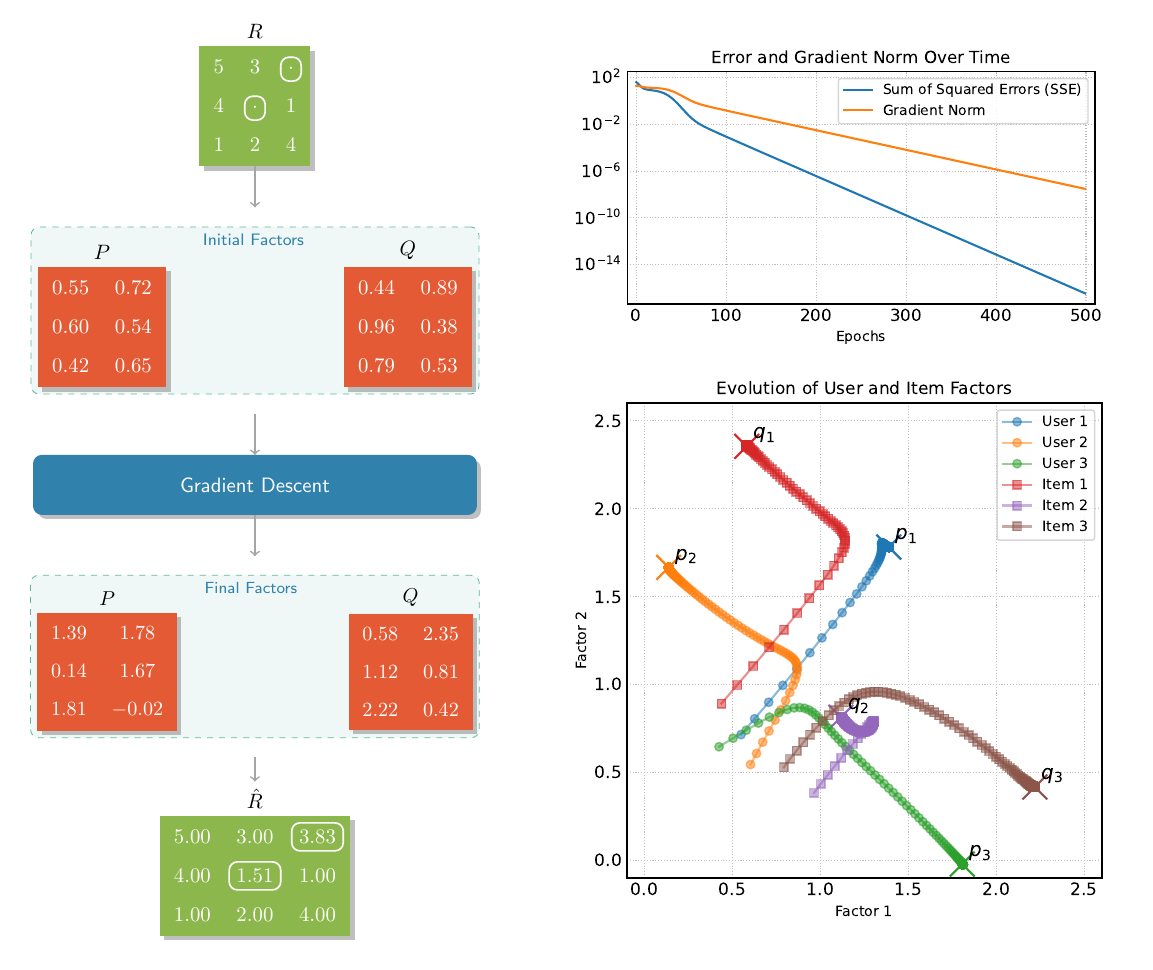}
		\caption{An example of using gradient descent to fit the matrix factorization model without regularization to data. Left: Starting from randomly generated values, gradient descent iteratively updates user and item factors until convergence. The final factors are used to predict the missing ratings. Right: The upper plot shows the evolution of the objective function (SSE) and the norm of the gradient throughout the optimization procedure. The bottom plot shows the evolution of user and item factors from their random initial positions to their final ones.}
		\label{fig:gd-example}
	\end{figure}

	In the experiment illustrated in Figure~\ref{fig:gd-example}, we use a small rating matrix consisting of three users and three items, with two missing ratings:$ r_{1,3} $ and $ r_{2,2} $. The goal is to predict these missing ratings by fitting a matrix factorization model using gradient descent. The process begins with randomly initializing the user and item factor vectors. The gradient descent algorithm, with learning rate $ \eta=0.01 $, then iteratively updates these factors to minimize the objective function. Our objective function does not include a regularization term and consists only of the sum of squared errors (SSE) between the predicted and actual ratings. The objective value and the gradient norm are recorded throughout the iterations to monitor the algorithm's progress. Upon convergence, the obtained factors are used to predict the missing ratings, which gives $\hat{r}_{1,3}=3.83$  and $\hat{r}_{2,2}=1.51$.
	
	Gradient descent requires computing the gradient using the entire training set, which can be computationally expensive for large datasets. To address this, Stochastic Gradient Descent (SGD) \cite{sgd} approximates the gradient using a mini-batch of the training set, which reduces computational costs. Several variants of SGD, including SGD with Nesterov Momentum \cite{nestrov}, AdaGrad \cite{adagrad}, RMSProp \cite{rmsprop}, and Adam \cite{adam}, are widely used in modern machine learning and deep learning models. These and other advanced optimization methods are discussed in detail in Section~\ref{sec:opt}.
	
	\section{Learning Data}
	\label{sec:learning-data}
	Collaborative filtering typically uses rating data as its primary input, which reflects user preferences through explicit feedback mechanisms. However, integrating contextual information can significantly enhance the accuracy and personalization of recommendations. In this section, we explore three key types of contextual data crucial for improving recommendation systems: implicit feedback, trust, and content data. Implicit feedback is indirectly gathered from user activities, such as browsing history and purchase records, and provides insights into user preferences without explicit ratings. Trust data, derived from social interactions and user endorsements, incorporates the relational trust dynamics into the recommendation process. Content data, including item descriptions and user profiles, offers detailed insights into the characteristics of items and users. Systems that combine content data with collaborative filtering techniques are known as hybrid models, combining the strengths of content-based and collaborative filtering approaches to improve recommendation accuracy and personalization.
	
	\subsection{Implicit Feedback}
	\label{sec:implicit-feedback}
	Implicit feedback is information collected indirectly from user interactions with items. This type of data includes browsing history, purchase records, or watch time, and it does not require explicit user actions such as ratings. Unlike explicit feedback, implicit feedback provides insights into user preferences through their behavior. It is especially valuable in systems where explicit ratings are sparse or absent, and it can enhance the quality of recommendations by providing additional insights into user preferences for items that were neither purchased nor explicitly rated. Existing approaches can be categorized into linear and nonlinear depending on how the implicit feedback component relates to the predicted ratings.
	
	\subsubsection{Linear Approaches}
	In linear approaches, the effect of implicit feedback is captured through latent factors that affect the predicted rating linearly. These often appear as additive terms that adjust the basic matrix factorization prediction.
	SVD++ \cite{svdpp} is such a model that integrates neighborhood models with matrix factorization to enhance recommendation systems by utilizing both explicit and implicit feedback. This approach merges the strengths of latent factor models and neighborhood-based methods within a unified framework, aiming to improve the accuracy of user preference predictions. The estimated rating given by user $u$ to item $i$ by SVD++ is expressed as:
	\begin{equation}
		\hat{r}_{ui} = \bar{r} + b_u + b_i + q_i^T \left( p_u + \frac{|N_u|^{-\frac{1}{2}}}{2} \sum_{j \in N_u} y_j \right) + \frac{|\mathcal{R}^k_{ui}|^{-\frac{1}{2}}}{2} \sum_{j \in \mathcal{R}^k_{ui}} w_{ij}(r_{uj} - b_{uj}) + \frac{|N^k_{ui}|^{-\frac{1}{2}}}{2} \sum_{j \in N^k_{ui}} c_{ij},
	\end{equation}
	where $\bar{r}$ is the global average rating across all users and items, $b_u$ is the bias term associated with user $u$, $b_i$ is the bias term associated with item $i$, $N_u$ is the set of items for which user $u$ showed an implicit preference, $y_j$ is the implicit feedback factor vector for item $j$, $\mathcal{R}^k_{ui} = \mathcal{R}_u \cap S^k_i$, where $\mathcal{R}_u$ is the set of items rated by user $u$ and $S^k_i$ the set of $k$ items most similar to $i$, $b_{uj}$ is the bias term for the interaction of user $u$ with item $j$, $w_{ij}$ is the weight of the influence of item $j$ on the rating of item $i$ by user $u$, $N^k_{ui} = N_k \cap S^k_i$, and $c_{ij}$ is a weight factor representing the relationship strength between items $i$ and $j$ for user $u$.
	
	Shi et al. \cite{shi2020user} proposed User Embedding for rating prediction in the SVD++ model \cite{koren2009matrix} (UE-SVD++). The UE-SVD++ model leverages user correlations from the user-item matrix and constructs a user embedding matrix to enhance prediction accuracy. Initially, the most favored users for each item are identified, specifically those whose ratings exceed 70\% of the highest rating. This restricted set of ratings is denoted by $\Tilde{\mathcal{R}}$. The list of users for each item is termed a context. Using $\Tilde{\mathcal{R}}$, User-wise Mutual Information (UMI) values are calculated as follows:
	\begin{equation}
		\label{eq:UESVD++}
		UMI(u, v) = \log \frac{p(u, v)}{p(u) p(v)},
	\end{equation}
	where $p(u, v)$ is the joint probability of users $u$ and $v$ appearing together in a context, and $p(u)$ and $p(v)$ are the probabilities of $u$ and $v$ appearing in any context, respectively. These probabilities are computed from the frequency of occurrences in $\Tilde{\mathcal{R}}$.
	A threshold $k$ is then applied to the UMI matrix to enhance its reliability:
	\begin{equation}
		\label{eq:threshold UMI}
		KMUI(u, v) = 
		\begin{cases} 
			UMI(u, v) & \text{if } UMI(u, v) \geq \log(k) \\
			0 & \text{otherwise}
		\end{cases}
	\end{equation}
	The embedding matrix $D$ is constructed based on KMUI as follows:
	\begin{equation}
		D_{uv} = 
		\begin{cases} 
			1 & \text{if } KMUI(u, v) > 0 \\
			0 & \text{otherwise}
		\end{cases}
	\end{equation}
	Define the set $D_u$ as the set of users $v$ such that $D_{uv} = 1$. The rating predicted by UE-SVD++ is an adaptation of the SVD++ rating \cite{svdpp} and is calculated as follows:
	\begin{equation}
		\hat{r}_{ui} = \bar{r} + b_u + b_i + q_i^T \left( p_u + |N_u|^{-\frac{1}{2}} \sum_{j \in N_u} y_j + |D_u|^{-\frac{1}{2}} \sum_{v \in D_u} z_v \right),
	\end{equation}
	where $\bar{r}$ is the global average rating across all users and items, $b_u$ is the bias term associated with user $u$, $b_i$ is the bias term associated with item $i$, $N_u$ is the set of items for which user $u$ showed an implicit preference, $y_j$ is the implicit feedback factor vector for item $j$, and $z_v$ is the factor vector for user-to-user dependency.
	
	\subsubsection{Nonlinear Approaches}
	Nonlinear methods capture user-item interactions and the relationship between implicit feedback and ratings through complex models, typically utilizing neural networks. For instance,  
	Zhang et al. \cite{zhang2021integrating} introduced Stacked Sparse Auto-encoder Recommender Systems (SSAERec). This model integrates a deep Stacked Sparse Autoencoder into matrix factorization, effectively learning the user-item matrix representation. It utilizes multiple layers of sparse autoencoders, a Sparse Autoencoders (SAE) variant, to enhance feature extraction. The middle-most layer is combined with Singular Value Decomposition++ (SVD++) \cite{koren2008factorization} to incorporate implicit feedback into the model and predict unknown ratings. For optimization, they employed the Adam algorithm \cite{adam} to optimize the Stacked Sparse Auto-encoder model and Stochastic Gradient Descent (SGD) \cite{funk2006netflix} to learn the rating prediction parameters.
	
\begin{figure}[!t]
		\centering
		\includegraphics[width=0.9\textwidth]{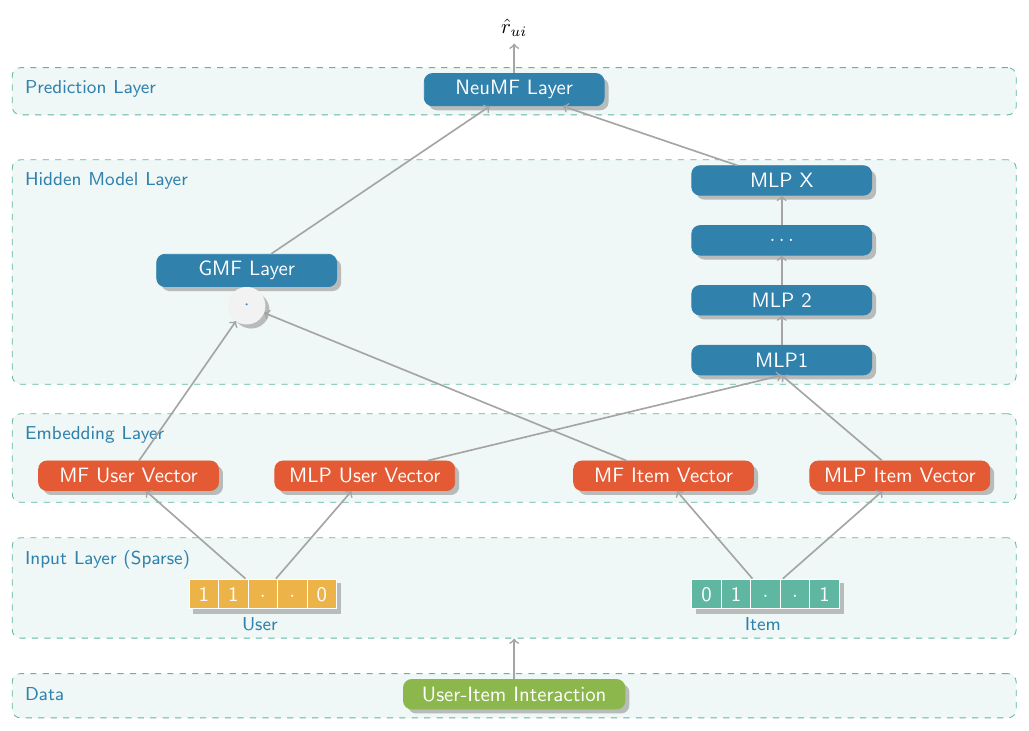}
		\caption{The architecture of the NCF model \cite{he2017neural}.}
		\label{fig:ncf}
	\end{figure}

	He et al. \cite{he2017neural} introduced the Neural Collaborative Filtering (NCF) model, which aims to enhance traditional collaborative filtering approaches by integrating neural networks. This model focuses on implicit feedback scenarios—such as clicks or views rather than explicit ratings—to learn user preferences more effectively. The NCF framework, shown in Figure~\ref{fig:ncf}, encompasses multiple instantiations, including Generalized Matrix Factorization (GMF), Multi-Layer Perceptron (MLP), and a combined model called NeuMF, which merges GMF and MLP to leverage both linearity and non-linearity in user-item interactions.
	The core idea is to replace the traditional dot product used in matrix factorization with a neural architecture that can learn an arbitrary function from user-item interactions. This approach allows NCF not only to mimic but also to extend the capabilities of matrix factorization models by introducing non-linearities through deep learning layers. Extensive testing on datasets like MovieLens and Pinterest showed that NCF significantly outperforms classical methods, particularly in handling the complex and often sparse data typically found in collaborative filtering systems.
	
\begin{figure}[!t]
		\centering
		\includegraphics[width=0.9\textwidth]{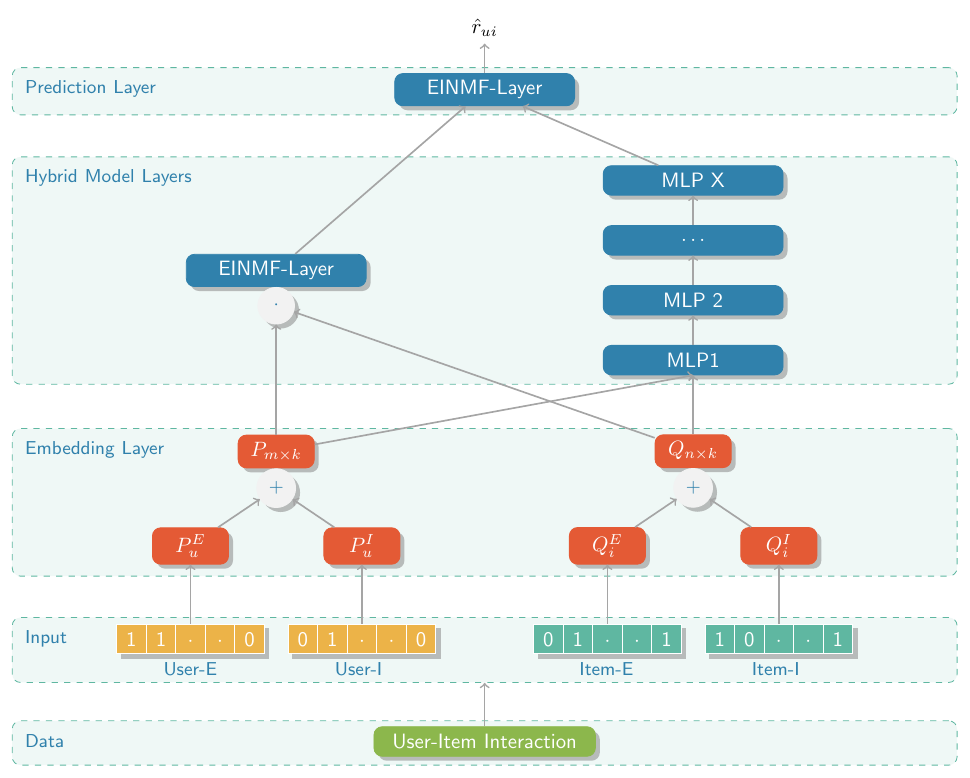}
		\caption{The architecture of the EINMF model \cite{liu2022neural}.}
		\label{fig:EINMF}
	\end{figure}

	Liu et al. \cite{liu2022neural} proposed a neural matrix factorization algorithm based on explicit-implicit feedback (EINMF), which learns both linear and nonlinear explicit-implicit feedback features for the user-item matrix. As illustrated in Figure~\ref{fig:EINMF}, the explicit rating matrix and the implicit feedback matrix are used as input for the model. Embedding with normal stochasticity was applied to obtain user and item explicit–implicit latent vectors. These complementary latent feature vectors are input for the hybrid model layer. In the hybrid model layer, matrix factorization was used to extract the linear preferences features with the dot product operation, and the Multilayer Perceptron (MLP) model was used to obtain the nonlinear preferences features for users and items. They use matrix factorization to extract linear preference features via the dot product operation, and a Multilayer Perceptron (MLP) to capture nonlinear preference features for users and items. Subsequently, outputs from the hybrid model are concatenated to predict the degree of user preferences. This process leads to the prediction of unknown item ratings, providing users with a personalized top-$N$ recommendation list. They introduced a new loss function tailored to explicit and implicit feedback and optimized the EINMF model parameters using forward and backward propagation.
	
	\subsection{Trust}
	\label{sec:trust}
	Collaborative filtering and recommendation systems benefit from incorporating trust data, which represents the level of confidence or reliance that a user places in the opinions or behaviors of others. This information is gathered from social interactions on platforms where users rate not just products or services, but also express their trust in other users’ recommendations. Trust data is typically found in social networks, user reviews, and interactive platforms where users can follow others or endorse their reviews. By integrating trust metrics into recommendation algorithms, we can leverage the social context of user interactions, which provides an additional layer of data that significantly enhances the accuracy and personalization of recommendations. This is especially useful in addressing challenges such as data sparsity and the cold start problem, where new users or items have insufficient interactions.
	
	TrustMF \cite{yang2016social} incorporates social trust into the collaborative filtering process to address the limitations of data sparsity and the cold start problem in recommendations. It utilizes matrix factorization techniques to integrate user rating data with social trust networks, mapping users into two low-dimensional spaces reflecting their roles as trusters and trustees. This dual representation captures the directional nature of trust, where a user's rating behavior is influenced by those they trust and, in turn, influences others. Like traditional matrix factorization, the authors assume that ratings can be expressed as the product of user and item factor vectors $p_u^T q_i$. The trust network is represented as a non-symmetric matrix $T$ where $T_{uv}$ indicates the trust level from user $u$ to user $v$, ranging from 0 (no trust) to 1 (complete trust). Each user $u$ is associated with two distinct latent feature vectors: a truster-specific vector $p_u$, the same as the vector used to compute ratings, and a trustee-specific vector $w_u$. These vectors capture the behaviors of trusting others and being trusted, respectively. The trust value $T_{uv}$ is modeled as the inner product $p_u^T w_v$, reflecting the directional nature of trust. The model fits the rating and trust matrices simultaneously by minimizing the following objective function:
	\begin{equation}
		J_{TrustMF} = \sum_{r_{ui} \in \mathcal{R}} \left(p_u^T q_i - r_{ui}\right)^2 + \lambda_T \sum_{(u,v) \in \Psi} \left(p_u^T w_v - T_{uv}\right)^2 + \lambda \left(\sum_{u=1}^{n} \|p_u\|_2^2 + \sum_{i=1}^{m} \|q_i\|_2^2 
		+ \sum_{u=1}^{n} \|w_u\|_2^2\right),
	\end{equation}
	where $\Psi$ denotes the set of observed trust relations, $\lambda_T$ is the weight given to the trust term, and $\lambda$ is the regularization coefficient. This approach has shown enhanced performance on the Epinions dataset, outperforming traditional collaborative filtering and other trust-aware methods, significantly improving the quality of recommendations.
	
	TrustSVD \cite{guo2015trustsvd} is an extension of the SVD++ algorithm that takes into account both explicit and implicit influences from user-item ratings and user-user trust. The explicit influence consists of numerical ratings and trust scores, while the implicit influence involves observing which users rated which items and who trusts whom. More precisely, the rating model of TrustSVD is given by:
	\begin{equation}
		\hat{r}_{ui} = \bar{r} + b_u + b_i + q_i^T \left(p_u + \frac{1}{|\mathcal{R}_u|} \sum_{j \in \mathcal{R}_u} y_j + \frac{1}{|T_u|} \sum_{v \in T_u} s_v\right),
	\end{equation}
	where $\bar{r}$ is the global average rating, $b_u$ is the bias of user $u$, $b_i$ is the bias of item $i$, $\mathcal{R}_u$ is the set of items rated by user $u$, $y_j$ is the implicit feedback for item $j$, $T_u$ is the set of users trusted by user $u$, and $s_v$ is the implicit trust feedback for user $v$. The authors analyzed the performance of their approach on several datasets and concluded that trust information is sparse but complements rating information, leading to improved recommendation accuracy.
	
	Incorporating social trust information as an additional data source has proven effective in improving prediction accuracy while reducing computational costs and enhancing convergence speed. Parvin et al. \cite{parvina2018efficient} proposed a social regularization method called Trust Regularized Single-element based Non-Negative Matrix Factorization (TrustRSNMF). The TrustRSNMF model extends the Regularized Single-element based Non-Negative Matrix Factorization (RSNMF) \cite{luo2015nonnegative} model by considering users' trust statements in the recommendation process. Users' trust statements were incorporated into the non-negative matrix factorization to improve prediction accuracy.
	
	\subsection{Content Data}
	\label{sec:content-data}
\begin{figure}[!t]
		\centering
		\includegraphics[width=0.8\textwidth]{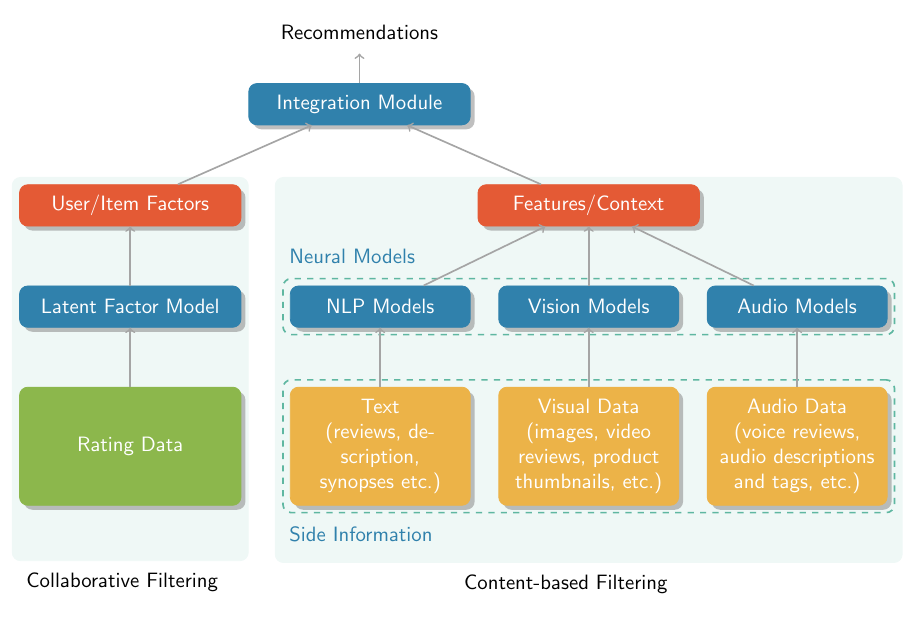}
		\caption{A hybrid recommendation system combines collaborative and content-based filtering methods to improve prediction accuracy. It leverages diverse data types such as textual, visual, and audio through neural models to extract relevant features. The extracted features are then combined with user/item factors to produce the final recommendations.}
		\label{fig:hybrid}
	\end{figure}

	Incorporating content data, such as item descriptions and user profiles, into rating data gives rise to hybrid models. This integration leverages additional information such as item descriptions, user profiles, and textual content commonly found in e-commerce product reviews or movie review sites. For instance, item descriptions in online shopping platforms or movie synopses in streaming services provide rich, descriptive data. Additionally, the availability of textual and sometimes imagery data creates an ideal scenario for employing neural networks, particularly deep learning models, to extract pertinent features that enhance recommendation accuracy. As illustrated in Figure~\ref{fig:hybrid}, most hybrid models combine a neural model, which processes textual or visual content, with a collaborative filtering framework. The features extracted by the neural model are then integrated with the outputs from the collaborative filtering model to produce more precise and personalized recommendations.
	
	Kim et al. \cite{kim2016convolutional} proposed a context-aware recommendation model called Convolutional Matrix Factorization (ConvMF), which integrates a Convolutional Neural Network (CNN) \cite{lecun1998gradient} into Probabilistic Matrix Factorization (PMF) \cite{mnih2007probabilistic}. Initially, a CNN analyzes documents' contextual information and generates latent document vectors. These vectors are then integrated with the PMF model to predict unknown ratings. Consequently, CNN provides a deep representation of documents, enhancing the accuracy of rating predictions.
	
	They applied maximum a posteriori (MAP) \cite{mnih2007probabilistic} estimation to optimize the latent model variables for users and items, as well as the weights and biases of the CNN. Three datasets were used for testing: MovieLens-1M, MovieLens-10M, and Amazon, with RMSE as the evaluation metric. The ConvMF model was compared with Probabilistic Matrix Factorization (PMF), Collaborative Deep Learning (CDL) \cite{wang2015collaborative}, and ConvMF with a pre-trained word embedding model (ConvMF+). The results demonstrated that both ConvMF and ConvMF+ significantly enhanced prediction accuracy compared to PMF across all datasets. 
	
	Mohd et al. \cite{mohd2021word} analyzed word documents using Long Short-Term Memory (LSTM) networks to transform product review documents into a 2D latent space vector. They then integrated this vector with Probabilistic Matrix Factorization (PMF) \cite{mnih2007probabilistic}, known for effectively generating rating predictions in large datasets and robustly handling imbalanced data.
	
	The first layer of LSTM-PMF is responsible for collecting the datasets. The second layer employs the Natural Language Toolkit (NLTK) for preprocessing and Global Vector (GloVe) for word embedding \cite{pennington2014glove}. The third layer uses LSTM to transform the product review documents into a 2D vector space. PMF integrates the item and user latent vectors in the fourth layer to generate a rating prediction. The final layer utilizes the Root Mean Square Error (RMSE) \cite{herlocker2004evaluating} metric to evaluate prediction accuracy. Learning and weight variables are optimized using Maximum A Posteriori distribution (MAP) \cite{mnih2007probabilistic} and the back-propagation algorithm. In their experiments with the MovieLens-1M and MovieLens-10M datasets, the proposed LSTM-PMF model was compared with traditional PMF and a Convolutional Neural Network-based PMF (CNN-PMF) \cite{kim2016convolutional}, using RMSE as the performance measure. 
	
	Sun et al. \cite{sun2020joint} introduced a probabilistic framework, Joint Matrix Factorization (JMF), incorporating user and item latent information alongside additional side information from product reviews. Differing from Mohd et al. \cite{mohd2021word}, they employed a modified Long Short-Term Memory (LSTM) \cite{schmidhuber1997long}, precisely Bidirectional Long Short-Term Memory (BLSTM), to capture information from both directions of a sequence.
	The JMF model comprises three components:
	\begin{enumerate}
		\item A Multilayer-crossing Factorization Machine (MFM), a variation of the Factorization Machine (FM) \cite{guo2017deepfm, rendle2010factorization} designed for improved computational efficiency and noise reduction. This component extracts latent user factors based on behavioral data.
		\item The BLSTM, which processes item latent factors from document sequences.
		\item These components are integrated with Probabilistic Matrix Factorization (PMF) to generate predictive ratings.
	\end{enumerate}
	Five datasets were utilized for their experiments: MovieLens-100k, MovieLens-1M, MovieLens-10M, Amazon Music, and Amazon Baby. The Root Mean Square Error (RMSE) \cite{herlocker2004evaluating} was employed as the evaluation metric.
	
	Ong et al. \cite{ong2021neural} explored linear and nonlinear user and item representations. They introduced the Neural Matrix Factorization++ (NeuMF++) model, an enhanced version of NeuMF \cite{he2017neural}. NeuMF++ incorporates side information using Stacked Denoising Autoencoders (SDAE) \cite{strub2015collaborative}, effectively capturing latent features.
	
	The NeuMF++ model consists of two main components: feature extraction and neural collaborative filtering. Initially, SDAEs process each user and item feature to derive latent representations. Subsequently, high-level features are extracted from the middle-most layer. NeuMF++ integrates Generalized Matrix Factorization (GMF++) to address linearity and a Multilayer Perceptron (MLP++) to manage nonlinearity in the neural collaborative filtering stage. The features extracted from SDAEs are treated as embeddings, concatenated, and fed into both GMF++ and MLP++. The outputs from GMF++ and MLP++ are then concatenated into a single MLP layer to predict unknown ratings. The Adam optimizer \cite{adam} is employed to refine the NeuMF++ model parameters. In their experiments using the MovieLens-1M dataset, the NeuMF++ model was benchmarked against the original NeuMF model using RMSE.
	
	Zhang et al. \cite{zhang2017autosvd++} introduced a new hybrid model that integrates a Contractive Auto-encoder (CAE) \cite{salah2011contractive} with Biased Singular Value Decomposition (BSVD) \cite{koren2008factorization}, termed AutoSVD. The CAE extracts complex, nonlinear feature representations of item information, which are then integrated into SVD to enhance learning and prediction of unknown ratings. They extended this approach with AutoSVD++, incorporating implicit feedback, side information, and the user-item rating matrix to address data sparsity better. Experiments were conducted across three datasets: MovieLens-100k, MovieLens-1M, and MovieTweetings. AutoSVD and AutoSVD++ were compared with SVD++ \cite{koren2009matrix} and Biased SVD \cite{koren2009matrix} using RMSE. 
	
	Nassar et al. \cite{nassar2020multi} suggested a multi-criteria collaborative filtering recommender system by integrating a Deep Neural Network (DNN) \cite{hinton2012deep} and Matrix Factorization (MF) \cite{koren2009matrix}. This integration aims to harness the non-linearity of DNN alongside the linearity of MF to improve rating predictions. The proposed model is an evolution of their previous work, the Deep Multi-Criteria Collaborative Filtering (DMCCF) \cite{nassar2020novel} model, differing primarily in its use of MF in conjunction with a DNN to predict criteria-specific ratings, whereas the original DMCCF utilized only DNN.
	
	The model operates in two phases. Initially, user and item features are input into a combined model of a DNN and MF to predict the criteria ratings. Subsequently, these criteria ratings are used as inputs for another DNN to predict the overall ratings. Both model components are optimized using the Adam optimizer \cite{adam} and the MAE loss function.
	
	Experiments conducted with the TripAdvisor and Movies datasets compared the performance of the proposed model against the DMCCF. The evaluation metric used was the RMSE. The results indicate that the proposed model performs better on both datasets and significantly outperforms the DMCCF model.
	
	\subsection{Discussion}
	
	\begin{table}[!t]
		\centering
		\caption{Summary of experimental results from surveyed papers using various additional learning data types.}
		\label{tab:learning-data}
		\begin{tabular}{p{3cm}m{4cm}m{3.5cm}m{6cm}}
			\toprule
			\textbf{Reference}                  & \textbf{Type of Additional Data}                                                                & \textbf{Datasets Used}                         & \textbf{Results}                                                                                                          \\ \midrule
			SVD++ \cite{svdpp}                  & Implicit feedback                                                  & Netflix                                        & RMSE: 0.887                                                                                                               \\ \midrule
			UE-SVD++ \cite{shi2020user}         & Implicit feedback                                                & MovieLens-100k, Epinions, FilmTrust, EachMovie & MovieLens-100k: RMSE 0.942, \newline Epinions: RMSE 1.058,\newline  FilmTrust: RMSE 0.802,\newline  EachMovie: RMSE 0.257 \\ \midrule
			SSAERec \cite{zhang2021integrating} & Implicit feedback                                           & Ciao, MovieLens-100k, MovieLens-1M             & MovieLens-100k: RMSE 0.902,\newline  MovieLens-1M: RMSE 0.848                                                             \\ \midrule
			NCF \cite{he2017neural}             & Implicit feedback                                      & MovieLens, Pinterest                           & MovieLens-1M: HR@10: 0.69, NDCG@10: 0.42 (8 factors) \newline HR@10: 0.87, NDCG@10: 0.55 (8 factors) methods                                                                               \\ \midrule
			EINMF \cite{liu2022neural}          & Implicit feedback & MovieLens-100K, MovieLens-1M                   & MovieLens-100K: HR 0.698, NDCG 0.316; MovieLens-1M: HR 0.654, NDCG 0.283                                                  \\ \midrule
			TrustMF \cite{yang2016social}         & Trust                            & Epinions                                       & Epinions: RMSE 1.059 \\ \midrule
			TrustSVD \cite{guo2015trustsvd} & Trust                            & Epinions, FilmTrust, Flixster, Ciao             & Epinions: RMSE 1.044, \newline FilmTrust: RMSE 0.787, \newline Flixster: RMSE 0.950, \newline Ciao: RMSE 0.956 \\ \midrule
			TrustRSNMF \cite{parvina2018efficient} & Trust                            & Epinions, FilmTrust                            & FilmTrust: MAE 0.603, RMSE 0.753, \newline Epinions: MAE 0.843, RMSE 1.052 \\ \midrule 
			Kim et al. \cite{kim2016convolutional} & Content data                        & MovieLens-1M, MovieLens-10M, Amazon            & MovieLens-1M: RMSE 0.853, \newline MovieLens-10M: RMSE 0.793, \newline Amazon: RMSE 1.128 \\ \midrule
			Mohd et al. \cite{mohd2021word}     & Content data                        & MovieLens-1M, MovieLens-10M                    & MovieLens-1M: RMSE 0.841, \newline MovieLens-10M: RMSE 0.790                                                               \\ \midrule
			Sun et al. \cite{sun2020joint}      & Content data                        & MovieLens-100k, MovieLens-1M, MovieLens-10M, Amazon Music, Amazon Baby & MovieLens-10M: RMSE 0.782, \newline MovieLens-1M: RMSE 0.841, \newline MovieLens-100k: RMSE 0.902           \\ \midrule
			Ong et al. \cite{ong2021neural}     & Content data                        & MovieLens-1M                                   & RMSE: 0.868                                                                              \\ \midrule
			Zhang et al. \cite{zhang2017autosvd++} & Content data                        & MovieLens-100k, MovieLens-1M, MovieTweetings   & MovieLens-100k: RMSE 0.901, \newline AutoSVD++: RMSE 0.904, \newline MovieLens-1M: RMSE 0.848 (AutoSVD++), \newline 0.864 (AutoSVD)                                \\ \midrule
			Nassar et al. \cite{nassar2020multi} & Content data                        & TripAdvisor                            & TripAdvisor: RMSE 0.743                                                                         \\ \bottomrule
		\end{tabular}
	\end{table}
	
	The section on learning data focuses on improving collaborative filtering by including contextual information. Three key types of contextual data are explored: implicit feedback, trust, and content data:
	\begin{itemize}
		\item Implicit feedback is gathered indirectly from user interactions like browsing history and purchase records, providing insights into user preferences without requiring explicit ratings. This data type is particularly valuable in scenarios where explicit ratings are sparse or unavailable, as it enhances the quality of recommendations by revealing user interests through their behavior.
		
		\item Trust data is derived from social interactions and user endorsements, integrating relational trust dynamics into the recommendation process. By incorporating trust data, recommendation systems can leverage the social context of user interactions, improving the accuracy and personalization of recommendations. This approach is especially useful for addressing data sparsity and the cold start problem.
		
		\item Content data includes information about items and users, such as item descriptions and user profiles. This data type enriches the recommendation process by providing additional context and characteristics of the items and users involved. Hybrid models, which combine content data with collaborative filtering techniques, leverage the strengths of both content-based and collaborative filtering approaches. This combination results in more precise and personalized recommendations, as neural networks and deep learning models can extract rich features from textual and imagery data.
	\end{itemize}
	Table~\ref{tab:learning-data} provides a summary of the main experimental results from surveyed papers using these three types of learning data. These research areas remain significant and continue to offer valuable insights and improvements for recommendation systems. There are also new research directions where integrating sophisticated data types to enhance the accuracy and personalization of recommendations promises to advance the field further.
	
	Multimodal data integration, which combines text, images, and audio, leverages deep learning for effective processing \cite{LiuQidong2023MRSA, GengShijie2023VTMF}. This approach helps create richer and more contextually aware recommendations by utilizing diverse data types.
	
	Context-aware recommendations use information like time and location to provide timely suggestions \cite{hariri2019context, CasilloMario2023Crsa, JeongSoo-Yeon2022DLCR}. By adapting to the user's current situation, these systems can deliver more relevant and personalized content.
	
	Physiological and behavioral signals, such as EEG and gaze tracking, capture deeper user insights for personalized recommendations \cite{PandaDebadrita2024AEns, sulikowski2021gaze, Castagnos10, SariJuniNurma2021PRBo, duchowski2007eye, YousefianJaziSaba2021Aemr}. These signals provide valuable information about users' cognitive and emotional states, allowing recommendation systems to tailor suggestions based on real-time user feedback and interactions.
	
	\section{Model}
	\label{sec:model}
	This section covers advanced modeling techniques that can significantly improve the performance of recommendation engines. We will discuss probabilistic models that use statistical methods to handle sparse and large-scale data, weighted models that assign varied importance to different factors, kernelized and nonlinear models that capture complex, nonlinear relationships within the data, and Graph Neural Networks (GNN), which leverage the relational structure of data to enhance the accuracy and personalization of recommendations. Each category represents an improvement over traditional matrix factorization methods, incorporating innovative strategies such as probabilistic inference, dynamic weighting, and high-dimensional data transformations to provide more accurate and personalized user-item recommendations.
	
	\subsection{Probabilistic Models}
	\label{sec:probabilistic-models}
	Probabilistic Matrix Factorization (PMF) \cite{mnih2007probabilistic} tackles the challenges of handling large-scale, sparse, and imbalanced datasets like those encountered in the Netflix Prize competition. Traditional latent factor models often struggle with such datasets due to their inability to manage missing data effectively and the computational demands posed by large-scale data. PMF addresses these issues by scaling linearly with the number of observations and implementing probabilistic linearity to handle data sparsity efficiently. The model asserts that user preferences can be expressed as a product of two lower-rank matrices, one for users and one for items, incorporating Gaussian noise to model the variability in user ratings. This approach not only provides more robust handling of sparse data but also improves prediction accuracy for users with few ratings by using a constrained version of PMF, which assumes users rating similar sets of movies have similar tastes. 
	
	PMF assumes a prior distribution on the latent factor vectors for both users and items to regularize these parameters:
	\begin{equation}
		p\left(P|\sigma_u^2\right) = \prod_{u=1}^{n} \mathcal{N}\left(p_u | 0, \sigma_U^2 \mathbf{I}\right),
		\quad
		p\left(Q|\sigma_i^2\right) = \prod_{i=1}^{m} \mathcal{N}\left(q_i | 0, \sigma_I^2 \mathbf{I}\right),
	\end{equation}
	where $\mathcal{N}$ is the normal distribution, $\sigma_U^2$ and $\sigma_I^2$ are the variances of these priors, and $ \mathbf{I} $ is the identity matrix. The probability of observing the set of ratings $ \mathcal{R} $ given the latent factors $ P, Q $ and noise variance $ \sigma^2 $ is defined as:
	\begin{equation}
		\label{eq:pmf-p}
		p\left(\mathcal{R} |P, Q, \sigma^2\right) = \prod_{r_{ui} \in \mathcal{R}} \mathcal{N}\left(r_{ui} | p_u^T q_i, \sigma^2\right),
	\end{equation}
	where $p_u$ and $q_i$ are the latent feature vectors for users and items. The optimization of this model involves minimizing a loss function that balances the reconstruction error and regularization terms to prevent overfitting, tailored through hyperparameters that control model complexity. By treating $ \sigma_U $, $ \sigma_I $ and $ \sigma $ as hyperparameters, maximizing the log-likelihood of \eqref{eq:pmf-p} amounts to minimizing:
	\begin{equation}
		J_{PMF} = \sum_{r_{ui} \in \mathcal{R}} \left( r_{ui} - p_u^T q_i \right)^2 + \lambda_U \sum_{u=1}^n \| p_u \|^2_2 + \lambda_I \sum_{i=1}^m \| q_i \|^2_2,
	\end{equation}
	where $ \lambda_U =  \sigma^2/\sigma_U^2$ and $ \lambda_I =  \sigma^2/\sigma_I^2$. The effectiveness of PMF and its extensions, such as constrained PMF and PMF with adaptive priors, was demonstrated through extensive experiments on the Netflix dataset. The combined use of PMF models and Restricted Boltzmann Machines led to a performance improvement of nearly 7\% over Netflix's algorithm, achieving an error rate of 0.886.
	
	\subsection{Weighted Models}
	\label{sec:weighted-models}
	Traditional latent factor models often assume that latent factors are weighted equally, which may not always be a reasonable assumption. This motivated Chen et al. \cite{chen2017weighted} to develop the Weighted-Singular Value Decomposition (WSVD) model, which assigns different weight values to latent factors to explain their importance. They used the SVD model \cite{yuan2019singular} to generate latent factors and then assigned a weight to each latent factor:
	\begin{equation}
		\label{eq:WSVD}
		\hat{r}_{ui}=\bar{r}+b_{u}+b_{i}+(w \odot p_{u} )^{T} \cdot q_{i},
	\end{equation}
	where $\bar{r}$ refers to the average of all ratings, $b_{u}$ and $b_{i}$ are the user bias for user $u$ and the item bias for item $i$, respectively. The vector $w$ contains the weights of the latent factors, and $p_{u}$ is the vector of user $u$'s latent factors. The operator $\odot$ denotes the Hadamard product between vectors $w$ and $p_{u}$, and $q_{i}$ is the vector of item $i$'s latent factors. Stochastic Gradient Descent (SGD) is used for optimization during the model's training. 
	
	In their experiments, they used five datasets: MovieLens-100k, MovieLens-1M, MovieLens-10M, FilmTrust, and Epinions. The WSVD model was compared with the SVD and SVD++ \cite{koren2009matrix}. WSVD achieved an RMSE of 0.943 on MovieLens-100k, 0.992 on MovieLens-1M, 0.947 on MovieLens-10M, and 1.093 on FilmTrust, outperforming baseline methods for all datasets.
	
	Gu et al. \cite{gu2020robust} also proposed a weighted SVD model (wSVD), which, unlike the model in \cite{chen2017weighted}, assigns weights to each rating in the rating matrix to reduce the effects of noise and unreliable ratings. This approach applies the SVD model to the original matrix to calculate an entrywise absolute error $e_{ui}$ as follows:
	
	\begin{equation}
		\label{eq:absoluterror}
		e_{ui} = \left| r_{ui}-p_{u}^{T}q_{i}\right|, r_{ui} \in \mathcal{R},
	\end{equation}
	
	where $r_{ui}$ is the rating given by user $u$ on item $i$, $p_{u}$ and $q_{i}$ are the user vector and the item vector, respectively. A lower weight is then assigned to entries with a high absolute error:
	
	\begin{equation}
		\label{eq:weight}
		w_{ui} =\phi (e_{ui}),
	\end{equation}
	
	where $\phi$ is a non-increasing mapping. After obtaining the weights, the model is obtained by minimizing the following objective function:
	\begin{equation}
		\label{eq:wSVD}
		J_{wSVD} =\sum_{r_{ui} \in \mathcal{R}} w_{ui} \left(r_{ui}-\bar{r} -b_u-b_i-p_u^Tq_i\right)^2 
		+\lambda \left(\sum_{u=1}^{n} \left(\|p_u\|_2^2 + b_u^2\right)+ \sum_{i=1}^{m} \left(\|q_i\|_2^2 + b_i^2 \right)\right),
	\end{equation}
	where $b_{u}$ and $b_i$ are the user bias for user $u$ and item bias for item $i$, respectively. Gu et al. \cite{gu2020robust} also suggested an initial guess to speed up the optimization process by employing a direct sparse SVD solver. They performed experiments on MovieLens-100k, MovieLens-1M, and MovieTweetings, using the Root Mean Square Error (RMSE) \cite{herlocker2004evaluating} metric to measure the model's performance. The proposed user-based weight (wSVDu) and user-item-based weight (wSVDui) were compared with Biased SVD (BSVD) \cite{koren2009matrix} and SVD++. The best result was achieved by the user-item-based weight (wSVDui) model with RMSE 0.839 for MovieLens-100k, 0.801 for MovieLens-1M, and 1.317 for MovieTweetings. The results showed that the user-item-based weight (wSVDui) model outperformed other models in all cases.
	
	\subsection{Kernelized and Nonlinear Models}
	\label{sec:kernelized-and-nonlinear-models}
\begin{figure}[!t]
		\centering
		\includegraphics[width=0.5\textwidth]{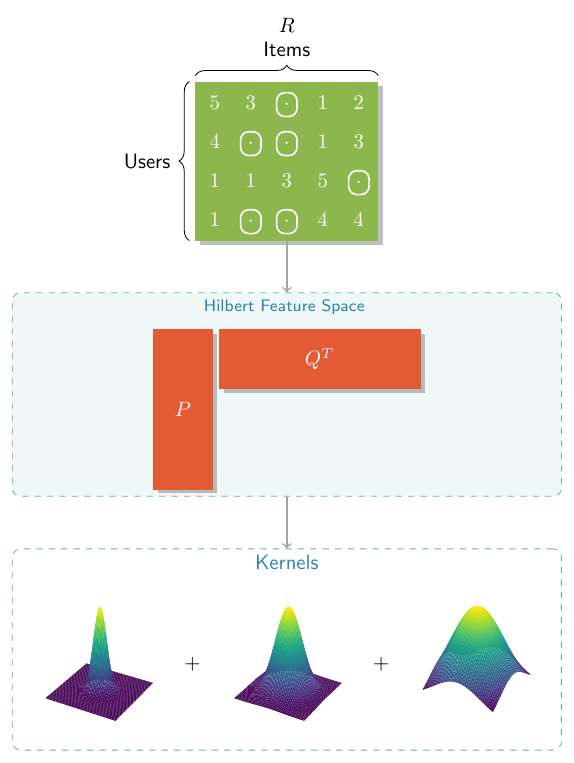}
		\caption{Kernelized low-rank matrix factorization \cite{liu2016kernelized} enhances prediction accuracy by embedding latent factor matrices $ P $ and $ Q $ into a high-dimensional Hilbert feature space using kernel functions. This non-linear reconstruction of the rating matrix $ R $ in its original space is achieved through the product of these transformed matrices.}
		\label{fig:kmf}
	\end{figure}

	Matrix factorization typically assumes data are distributed on a linear hyperplane, which is not always true. Liu et al. \cite{liu2016kernelized} proposed Kernel Matrix Factorization (KMF), integrating matrix factorization with kernel methods \cite{scholkopf2002learning}. KMF embeds the latent factor matrices in a higher-dimensional feature space, as shown in Figure~\ref{fig:kmf}, to learn the non-linear correlations of the ratings in the original space.
	
	Kernel methods implicitly embed data into a high-dimensional, possibly infinite-dimensional, space using a kernel. Let $\mathcal{X}$ denote the original data space, and $\mathcal{H}$ the high-dimensional feature space, which has the structure of a Hilbert space and is sometimes referred to as the kernel space. Denote by $\phi: \mathcal{X} \to \mathcal{H}$ the embedding map that associates each data point $x$ with its embedding $\phi(x)$. The mapping $\phi$ is generally only implicitly defined and observed only through the inner product $\phi(x)^T \phi(x') \in \mathbb{R}$. A widely used choice of kernel is the Gaussian kernel:
	\begin{equation}
		K(x, x') = \phi(x)^T \phi(x') = \exp\left(-\frac{\| x-x'\|^2}{2 \sigma^2}\right),
	\end{equation}
	where $\sigma^2$ is the bandwidth parameter. We assume that $\mathcal{X}= \mathbb{R}^d$, where $d$ is a hyperparameter of the model. Each user $u$ is represented by a vector $a_u \in \mathbb{R}^d$ and each item $i$ by a vector $b_i \in \mathbb{R}^d$. Let $k$ denote the number of factors, that is, the dimensionality of the latent factor space ($k$ is generally different from $d$). Start by randomly selecting $k$ vectors $d_1, \ldots, d_k \in \mathbb{R}^d$. We refer to these vectors as the dictionary vectors. We assume that we can write $\phi(a_u)$ as a linear combination of the images of the dictionary vectors:
	\begin{equation}
		\phi(a_u) = \sum_{j=1}^{k} p_{uj} \phi(d_j) = \Phi p_u,
	\end{equation}
	where $\Phi=\left(\phi(d_1), \ldots, \phi(d_k)\right)$, and $p_u=\left(p_{u1}, \ldots, p_{uk}\right)$ is the latent factor vector associated with user $u$. The same applies to items:
	\begin{equation}
		\phi(b_i) = \sum_{j=1}^{k} q_{ij} \phi(d_j) = \Phi q_i.
	\end{equation}
	The rating given by user $u$ to item $i$ is estimated as the inner product of $\phi(p_u)$ and $\phi(q_i)$ in the kernel space:
	\begin{equation}
		\hat{r}_{ui} = \phi(p_u)^T \phi(q_i) = (\Phi p_u)^T \Phi q_i = p_u^T (\Phi^T \Phi) q_i = p_u^T \mathbf{K} q_i.
	\end{equation}
	The matrix $\mathbf{K}$ is the Gram matrix associated with the dictionary vectors $\{d_i\}$, obtained by applying the kernel $K$ to each pair of dictionary vectors $(d_i, d_j)$:
	\begin{equation}
		\mathbf{K}_{ij} = K(d_i, d_j).
	\end{equation}
	The matrices $P$ and $Q$ are obtained by minimizing the following objective:
	\begin{equation}
		J_{KFM} = \sum_{r_{ui} \in \mathcal{R}} \left(r_{ui} - p_u^T \mathbf{K} q_i \right)^2 + \lambda \left( \sum_{u=1}^n \| p_u \|^2_2 + \sum_{i=1}^m \| q_i \|^2_2\right)
	\end{equation}
	
	Inspired by Multiple Kernel Learning (MKL) methods \cite{bach2004multiple, gonen2011multiple, lanckriet2004learning}, KMF extends to Multiple KMF (MKMF). MKMF combines multiple kernels, which learn a set of weights for each kernel function based on observed data in the rating matrix to improve prediction accuracy. The predicted rating is written as the weighted sum of multiple kernel inner products:
	\begin{equation}
		\hat{r}_{ui} = \sum_{j=1}^{l} w_j p_u^T \mathbf{K}_j q_i, \quad \text{where } \sum_{j=1}^l w_j = 1, w_j \geq 0, j=1, \ldots, l.
	\end{equation}
	The parameter $l$ is the number of kernels, $\mathbf{K}_j$ denotes the $j$-th kernel, and $w_j$ denotes the weight given to $\mathbf{K}_j$. The objective function for the case of multiple kernels is defined as:
	\begin{equation}
		J_{MKFM} = \sum_{r_{ui} \in \mathcal{R}} \left(r_{ui} - \sum_{j=1}^l w_j p_u^T \mathbf{K}_j q_i \right)^2 + \lambda \left( \sum_{u=1}^n \| p_u \|^2_2 + \sum_{i=1}^m \| q_i \|^2_2\right) + \lambda'  \sum_{j=1}^{l} w_j^2,
	\end{equation}
	where $\lambda$ and $\lambda'$ are regularization coefficients. The authors solve the minimization problem using an alternating algorithm, where first $Q$ is fixed and the best $P$ is found, then $P$ is fixed, and the best $Q$ is calculated until convergence. They performed experiments on several datasets: MovieLens, Flixster, Jester, Yahoo Music, ASSISTments, and Dating Agency, using the Root Mean Square Error (RMSE) \cite{herlocker2004evaluating} as the performance measure. They compared Kernel Matrix Factorization (KMF), which is based on a single kernel, with Multiple Kernel Matrix Factorization (MKMF), Matrix Factorization \cite{koren2009matrix}, and SVD \cite{koren2009matrix}. The results indicated that the MKMF model improved prediction accuracy compared with KML due to the use of multiple kernels. Additionally, MKMF outperformed both Matrix Factorization and SVD. The best RMSE results for MKMF were 0.816 on MovieLens, 0.815 on Flixster, 4.081 on Jester, and 18.503 on Yahoo Music.
	
	While the kernelized approaches, such as KMF, offer the advantage of embedding data into a higher-dimensional space through a non-linear mapping, they maintain a linear relationship with respect to the model parameters, allowing for efficient optimization. Lawrence et al. \cite{lawrence2009non} explored the potential of a nonlinear representation of latent factors to enhance prediction accuracy in matrix factorization. They proposed a nonlinear variant of Probabilistic Matrix Factorization (PMF) \cite{mnih2007probabilistic} by integrating the Gaussian Process Latent Variable Model (GP-LVM) \cite{seeger2004gaussian}. This nonlinear approach involved substituting the inner product in PMF with a Radial Basis Function (RBF) kernel \cite{han2012parameter}. The latent representations were optimized using Stochastic Gradient Descent (SGD) \cite{funk2006netflix}.
	
	In their experiments with the MovieLens-1M, MovieLens-10M, and EachMovie datasets, they compared the performance of the proposed nonlinear model to that of a conventional linear model using Root Mean Square Error (RMSE) \cite{herlocker2004evaluating} as one of the metrics. The RMSE values on MovieLens-1M were 0.875 for the nonlinear RBF model and 0.878 for the linear model. These results indicate that the nonlinear kernel slightly improved prediction accuracy compared to the linear model.
	
	\begin{figure}[!t]
		\centering
		\includegraphics[width=0.7\textwidth]{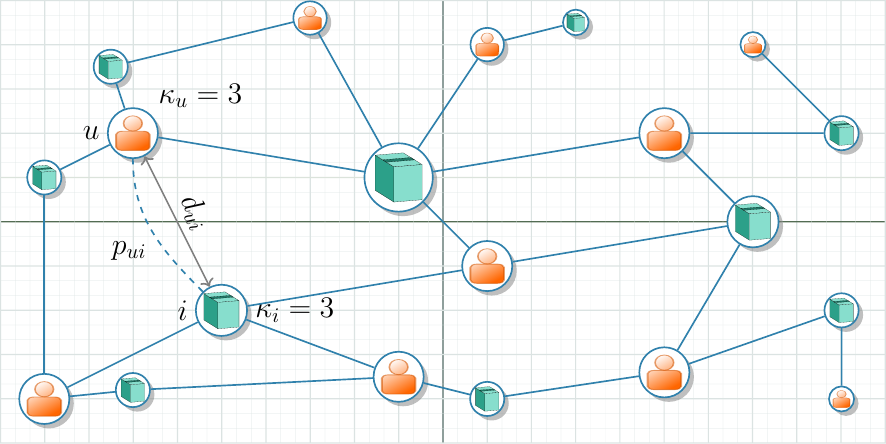}
		\caption{The SPHM model \cite{alhadlaq2022recommendation} fits the rating data to a similarity-popularity network model based on a hidden metric. The predicted rating from user $u$ to item $i$ is proportional to the probability of connection $p_{ui}$, which increases with the user degree $\kappa_u$ and item degree $\kappa_i$ and decreases with the distance $d_{ui}$ between them.}
		\label{fig:sphm}
	\end{figure}
	
	Authors in \cite{alhadlaq2022recommendation} proposed a novel method for recommender systems that leverages the structure of complex networks. This method models users and items as nodes within a network, using a similarity-popularity model to predict ratings. The primary goal is to address common challenges in recommender systems, such as data sparsity and the cold-start problem, by deriving insight from complex network models.
	
	Similarity-popularity models assume that two factors control node connectivity: their similarity and their popularity. More similar nodes tend to connect. Popularity, generally reflected by the node degree, is an innate property of the node that indicates its capacity to connect to other nodes. Popular nodes connect to other nodes even if they are highly dissimilar. The hidden metric space model is a similarity-popularity model where similarity between nodes is prescribed by an underlying hidden space metric, where similarity is the inverse of the distance.
	
	This approach fits a network model to the rating data, meaning that the model implicitly defines the network, but no actual edges are created (see Figure~\ref{fig:sphm}). The SPHM (Similarity-Popularity Hidden Metric) model proposed in \cite{alhadlaq2022recommendation} first scales the rating to the interval $ [p_{\min}, p_{\max}] \subset (0,1) $:
	\begin{equation}
		\label{eq:sphm-scale}
		\tilde{r}_{ui} = \phi(r_{ui}) = \frac{r_{ui} - r_{\min}}{r_{\max} - r_{\min}} (p_{\max} - p_{\min}) + p_{\min}.
	\end{equation}
	The probability of connections between the user node and item node gives the predicted scaled ratings in SPHM:
	\begin{equation}
		\label{eq:sphm-rating}
		\hat{\tilde{r}}_{ui} = \left(1 + \frac{d^2(p_u, q_i)}{\sqrt{\kappa_u \kappa_i}}\right)^{-1},
	\end{equation}
	where $ d(p_u, q_i) $ stands for the Euclidean distance between $ p_u $ and $ q_i $, and $ \kappa_u $ and $ \kappa_i $ are the degrees associated with the user $ u $ and item $ i $ nodes respectively. These are defined as:
	\begin{align}
		\kappa_u &= \bar{r}_u - r_{\min} + 1, \\
		\kappa_i &= \bar{r}_i - r_{\min} + 1,
	\end{align}
	where $ \bar{r}_u $ and $ \bar{r}_i $ are the average ratings of user $ u $ and item $ i $ respectively, and $ r_{\min} $ is the minimum possible rating in the system. The factors $ p_u $ and $ q_i $ are obtained by minimizing the following objective function:
	\begin{equation}
		J_{SPHM} = \sum_{\tilde{r}_{ui} \in \tilde{\mathcal{R}}} \left(\hat{\tilde{r}}_{ui} - \tilde{r}_{ui}\right)^2 + \lambda \left(\sum_{u=1}^n \| p_u \|^2_2 + \sum_{i=1}^m \| q_i \|^2_2\right),
	\end{equation}
	where $ \tilde{\mathcal{R}} $ is the set of scaled ratings as defined in Equation~\eqref{eq:sphm-scale}, and $ \hat{\tilde{r}}_{ui} $ is the predicted scaled rating defined in Equation~\eqref{eq:sphm-rating}. The authors use the conjugate gradient algorithm proposed in \cite{hager2005new} to solve this optimization problem. The predicted ratings are then obtained using the inverse of $ \phi $. SPHM's performance was demonstrated through an extensive experimental analysis conducted on 21 datasets against ItemKNN, SVD++, PMF, and BiasedMF.
	
	Neural models are highly adaptable and can be used not only in hybrid systems but also in purely collaborative filtering contexts. Their nonlinear modeling capabilities help improve the recommendation quality by capturing intricate patterns in user-item interactions that linear methods might overlook. For instance, Shang et al. \cite{shang2013extreme} proposed a novel nonlinear regression model that combines the Extreme Learning Machine (ELM) \cite{huang2004extreme} with Weighted Nonnegative Matrix Tri-Factorization (WNMTF) \cite{gu2010collaborative, ding2006orthogonal, seung2001algorithms}, named WNMTF Combined ELM for Collaborative Filtering (CELMCF) algorithm. WNMTF was employed as a preprocessing step to initialize unobserved ratings in the user-item matrix, effectively mitigating the data sparsity issue. Subsequently, the ELM algorithm, which includes a single hidden layer with several nodes, was used to generate recommendations. Their experiments utilized several datasets, including MovieLens, BookCrossing, Jester, and Tencent Weibo. They compared their model with Memory-based Collaborative Filtering (MemCF) \cite{su2009survey} and ELMCF, using the Mean Absolute Error (MAE) \cite{herlocker2004evaluating} as the evaluation metric. The results indicated that the proposed CELMCF model outperformed all other algorithms on these datasets. Specifically, CELMCF achieved the best MAE results of 0.653 on MovieLens, 0.602 on Jester, 0.521 on BookCrossing, and 0.105 on Tencent Weibo.
	
	\subsection{Graph Neural Networks}
	\label{sec:gnn}
	Graph neural networks (GNNs) are a type of neural network that is specialized in processing complex structured data, extending deep neural network applications to graph data structures \cite{WuZonghan2021ACSo}. These networks excel in tasks such as social network analysis, bioinformatics, and computer vision by utilizing the detailed relational information within graphs. Graph convolutional networks (GCNs), a prominent variant of GNNs, are particularly effective at learning graph representations by aggregating neighborhood features, similar to traditional convolution operations in neural networks \cite{KipfThomasN2017SCwG}. This approach adapts convolutional kernels from image processing to graph data, facilitating parameter sharing among nodes and simplifying the model, ultimately reducing training time \cite{NikolentzosGiannis2021GKAS}.
	
	\begin{figure}[!t]
		\centering
		\includegraphics[width=\textwidth]{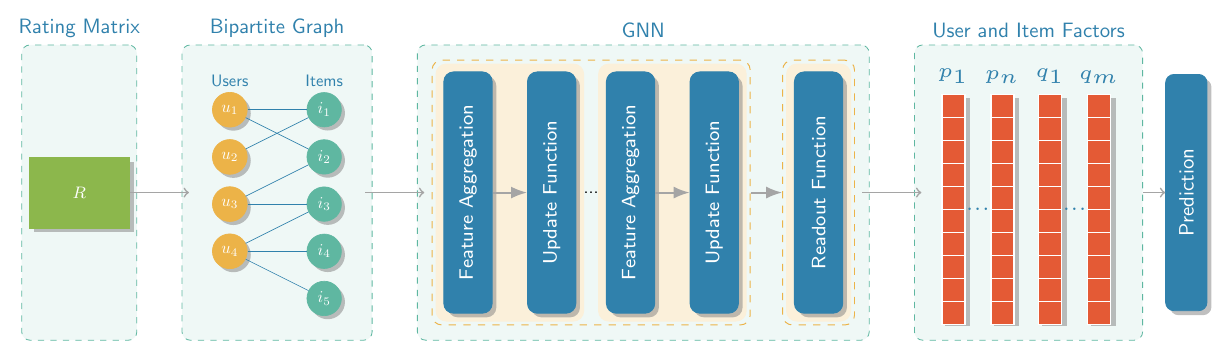}
		\caption{Using GNN for collaborative filtering \cite{wu2022graph}.}
		\label{fig:gnn}
	\end{figure}
	
	A GNN generally comprises multiple blocks in cascade, each performing two critical operations at each node of the graph \cite{wu2022graph}: feature aggregation, where information from the neighbors is collected, and node representation update, where the vector embedding of the node is updated based on the collected features. The last block of a GNN implements a readout function that uses the output of the previous blocks to generate the final node representations.
	
	As shown in Figure~\ref{fig:gnn}, to use a GNN for collaborative filtering, a graph is first constructed using the rating matrix. The graph contains two types of nodes: user and item nodes. The GNN then takes this graph as input, and the node embedding vectors are used as factors, which are then used to make predictions. For each of these steps, a variety of design choices are available, each influencing the effectiveness and efficiency of the network:
	\begin{itemize}
		\item \textbf{Graph construction}: Most works have applied GNN on user-item bipartite graphs. However, directly applying GNN on the original graph may not be effective or efficient due to the graph structure and computation cost \cite{berg2017graph,ChenLei2020RGBC,HeXiangnan2020LSaP,Jianing20,WangHaoyu2019BCFw,scc-10}. To address these issues, strategies such as adding edges between two-hop neighbors \cite{LiuZhiwei2021DGCF,SunJianing2019MCCF}, introducing virtual nodes \cite{xiang-23}, and sampling techniques \cite{rex-10,SunJianing2019MCCF} have been proposed. These strategies aim to enrich the original graph structure, improve expressiveness and computational efficiency. The choice of sampling strategy affects the performance of the model and requires further study.
		
		\item \textbf{Feature aggregation}: is a crucial part of information propagation for graph structures. Mean-pooling is a simple and popular aggregation operation \cite{berg2017graph, SunJianing2019MCCF, Qiaoyu20, ZhangMuhan2020IMCB}, but it may not be suitable when the importance of neighbors varies significantly. Other works use "degree normalization" to assign weights to nodes based on the graph structure \cite{ChenLei2020RGBC,TaoZhulin2023SLfM,wu-20}. Some methods use an attention mechanism to learn the weights of neighbors \cite{WangXiao2020MGCC, Jianxin20}.
		
		\item \textbf{Update function} is crucial for iterative information propagation in graph neural models. Existing methods can be classified into two categories based on whether they discard original node information or not. Some approaches only consider the information from the neighbors \cite{berg2017graph,HeXiangnan2020LSaP,Jianxin20}, while others combine the node information with its neighbors' information \cite{Jianing20,HeXiangnan2020LSaP,wu-20,ZhangMuhan2020IMCB}. The latter method usually involves concatenation functions with non-linear transformations, which allows for more complex feature interactions. However, some recent works simplify the update operation by removing non-linearities \cite{ChenLei2020RGBC,HeXiangnan2020LSaP}, which increases computational efficiency while retaining or even improving performance.
		
		\item \textbf{Readout function}: Different methods are used to generate the final representations of nodes used as the GNN output. The simplest and most common approach is to use the output of the last layer as the final representation \cite{berg2017graph, Qiaoyu20,rex-10}. Recent studies integrate the messages from different layers to take advantage of the connections expressed by the output of different layers \cite{ngcf19}. Mean-pooling, sum-pooling, weighted-pooling, and concatenation are examples of methods employed to integrate the messages.
	\end{itemize}
	
	GNNs have shown promising results in improving the recommendation quality of collaborative filtering. They capture complex, structured relationships in user-item interaction data and provide a flexible framework to encode direct and higher-order relationships. However, further research is needed to allow GNNs to handle dynamic graph data, increase their efficiency, and ensure their scalability \cite{wu2022graph}. 
	
	\subsection{Other Models}
	\label{sec:other-models}
	Researchers have considered the impact of time in latent factor models as user preferences change over time. Xiang et al. \cite{xiang2009time} integrated temporal dynamics into Regularized Singular Value Decomposition (RSVD) \cite{koren2008factorization} to enhance prediction accuracy. They identified four types of time effects within RSVD: user bias shifting, where a user may change ratings over time; item bias, indicating changes in the popularity of items; time bias, acknowledging shifts in society's interests and preferences over time; and user preference shifting, where a user may alter their opinion about some items:
	\begin{equation} 
		\label{eq:timeSVD} 
		\hat{r}_{ui} = \bar{r} + b_u + b_i + b_t + p_{u}^{T} q_i + x_{u}^{T} z_\tau + s_{i}^{T} y_\omega + \sum_{k} g_{uk} l_{ik} h_{\tau k}, 
	\end{equation}
	where $\bar{r}$ is the average of all known ratings, $b_u$ and $b_i$ are the user and item bias terms, respectively, $\tau$ represents the number of days since user $u$ joined the system, $x_u$ and $z_\tau$ are the latent factor vectors for user $u$ and time $\tau$, respectively, $\omega$ denotes the number of days since the first rating was assigned for item $i$, and $g_u$, $l_i$, and $h_\tau$ are three latent factor vectors for user $u$, item $i$, and time $\tau$, respectively.
	
	The authors utilized the MovieLens-1M and Netflix datasets, employing the RMSE evaluation metric for their experiments. The results demonstrated that TimeSVD improved prediction accuracy compared to the time-independent RSVD, achieving an RMSE of 0.836 on MovieLens-1M and 0.901 on the Netflix dataset.
	
	Some researchers have integrated clustering methods with latent factor models to enhance recommendation performance. Zarzour et al. \cite{zarzour2018new} combined Singular Value Decomposition (SVD) \cite{koren2009matrix} with the $k$-means algorithm \cite{o1999clustering} to cluster similar users and reduce dimensionality. Their approach uses the $k$-means algorithm to cluster users' ratings to determine the center-item rating matrix, which is then subjected to SVD to reduce dimensionality. Cosine similarity \cite{billsus2000user} is subsequently used to calculate similarities. For their experiments, they utilized the MovieLens-1M and MovieLens-10M datasets. They compared their model to traditional $k$-means using the RMSE. The results demonstrated that their approach outperformed $k$-means across all datasets, achieving the best RMSE of 0.620 on MovieLens-10M.
	
	Vozalis et al. \cite{vozalis2005applying} combined singular value decomposition (SVD) \cite{koren2009matrix} with item-based filtering to minimize the dimensionality of the user-item matrix. First, they applied the SVD on the user-item rating matrix to minimize dimensionality. Then, compute the similarity between the items using Adjacent Cosine Similarity \cite{sarwar2001item} and predict the unknown ratings.
	They used the MovieLens-100k dataset and the Mean Absolute Error to measure the model's performance. They compared the proposed model with basic item-based filtering. The proposed model outperformed basic item-based filtering, with 0.797, compared to basic item-based filtering, with 0.840.
	
	\subsection{Discussion}
	
	\begin{table}[!t]
		\centering
		\caption{Summary of experimental results from surveyed papers using various models.}
		\label{tab:models}
		\begin{tabular}{m{3cm}m{4cm}m{3.5cm}m{6cm}}
			\toprule
			\textbf{Reference} & \textbf{Model Type} & \textbf{Datasets Used} & \textbf{Results} \\ \midrule
			PMF \cite{mnih2007probabilistic} & Probabilistic & Netflix & Netflix: Error rate 0.886 \\ \midrule
			WSVD \cite{chen2017weighted} & Weighted & MovieLens-100k, MovieLens-1M, MovieLens-10M, FilmTrust & MovieLens-100k: RMSE 0.943, \newline MovieLens-1M: RMSE 0.992, \newline MovieLens-10M: RMSE 0.947, \newline FilmTrust: RMSE 1.093 \\ \midrule
			wSVD \cite{gu2020robust} & Weighted & MovieLens-100k, MovieLens-1M, MovieTweetings & MovieLens-100k: RMSE 0.839, \newline MovieLens-1M: RMSE 0.801, \newline MovieTweetings: RMSE 1.317 \\ \midrule
			KMF \cite{liu2016kernelized} & Kernelized & MovieLens, Flixster, Jester, Yahoo Music & MovieLens: RMSE 0.816, \newline Flixster: RMSE 0.815, \newline Jester: RMSE 4.081, \newline Yahoo Music: RMSE 18.503\\ \midrule
			SPHM \cite{alhadlaq2022recommendation} & Nonlinear & AmazonDM, AmazonIV, Anime, Book-Crossing, CiaoDVD, Epinions, FilmTrust, Food.com, ML100K, ML1M, YahooMovies, YahooMusic & AmazonDM: RMSE 0.784, MAE 0.484 \newline AmazonIV: RMSE 1.077, MAE 0.780 \newline Anime: RMSE 1.138, MAE 0.863 \newline Book-Crossing: RMSE 3.416, MAE 2.767 \newline CiaoDVD: RMSE 0.982, MAE 0.752 \newline Epinions: RMSE 1.060, MAE 0.819 \newline FilmTrust: RMSE 0.791, MAE 0.613 \newline Food.com: RMSE 0.938, MAE 0.558 \newline ML100K: RMSE 0.912, MAE 0.724 \newline ML1M: RMSE 0.853, MAE 0.676 \newline YahooMovies: RMSE 2.941, MAE 2.200 \newline YahooMusic: RMSE 1.190, MAE 0.989 \\ \bottomrule
			CELMCF \cite{shang2013extreme} & Neural & MovieLens, BookCrossing, Jester, Tencent Weibo & MovieLens: MAE 0.653, \newline Jester: MAE 0.602, \newline BookCrossing: MAE 0.521, \newline Tencent Weibo: MAE 0.105 \\ \midrule
			TimeSVD \cite{xiang2009time} & Other & MovieLens-1M, Netflix & MovieLens-1M: RMSE 0.836, \newline Netflix: RMSE 0.901 \\ \midrule
			SVD+k-means \cite{zarzour2018new} & Other & MovieLens-10M & MovieLens-10M: RMSE 0.620 \\ \midrule
			SVD+Item-based \cite{vozalis2005applying} & Other & MovieLens-100k & MovieLens-100k: MAE 0.797 \\ \bottomrule
		\end{tabular}
	\end{table}
	
	This section explored advanced modeling techniques that are pivotal in recommendation systems. The models under discussion, including probabilistic, weighted, nonlinear, kernelized, and neural models, are designed to tackle the challenges of collaborative filtering. These models aim to enhance prediction accuracy, handle sparse data, address the cold-start problem, and capture intricate patterns in user-item interactions. The key experimental findings from the papers reviewed in this section are succinctly presented in Table ~ ef {tab:models}.
	
	Probabilistic models offer a robust theoretical framework for handling large-scale and sparse datasets. Models like Probabilistic Matrix Factorization (PMF) can incorporate prior knowledge through prior distributions, effectively managing uncertainty and variability in user ratings while providing a deeper understanding of user preferences and item characteristics.
	
	Nonlinear models, with their ability to capture complex patterns that linear models may overlook, present a fascinating challenge. However, these models also come with their share of difficulties, such as optimization issues and scalability concerns, which can make them less favorable for massive datasets. Yet, the potential they hold for uncovering hidden patterns and improving recommendation accuracy is a compelling motivation for further exploration.
	
	Kernelized models balance linear simplicity and nonlinear flexibility by introducing nonlinearities at the feature level through kernel functions. This allows them to maintain efficient optimization routines similar to linear models, making them useful for scenarios where traditional linear models fail to capture the underlying complexities of the data.
	
	While there is a trend towards more sophisticated nonlinear approaches, particularly those employing neural networks, most implementations focus on processing auxiliary data, such as textual information in hybrid models, rather than enhancing the collaborative filtering process. An exception to this trend is the use of Graph Neural Networks (GNNs), which leverage the relational structure of data intrinsic to user-item interactions in recommendation systems. However, GNNs are still in the early stages of development, with research focused on overcoming computational overheads to enhance their efficiency and scalability.
	
	A proposed link to network geometry opens the possibility of using advanced techniques developed for modeling complex networks, particularly hyperbolic geometry models, to probe the geometry of recommender system data. Embedding recommendation data into hyperbolic space can leverage geometric properties to enhance similarity measures, efficiently represent sparse data, and improve the scalability and accuracy of recommendations.
	
	While the field is gradually shifting towards more complex and powerful neural network models for collaborative filtering, significant research efforts are still required to make these models more efficient and scalable for real-world applications, further enhancing the capabilities of recommendation systems.
	
	The increasing complexity of recommendation models, especially with deep neural networks, has underscored the need for explainable recommendations. Recent trends focus on enhancing transparency and user trust by providing clear, interpretable reasons for recommendations. Techniques such as attention mechanisms, feature importance visualization, and natural language explanations are being integrated into recommendation algorithms \cite{ZhangYongfeng2020ERAS}. These advancements not only help users understand the decision-making process but also increase satisfaction and trust. They also enable developers to diagnose and mitigate biases and errors, leading to more reliable recommendation systems. Explainable recommendation models are an exciting future research direction with the potential to enhance user experience and system robustness significantly.
	
	\section{Learning Strategy}
	\label{sec:learning-strategy}
	This section covers advanced learning strategies designed to enhance the performance of recommendation systems. We explore a variety of approaches, including Self-Supervised Learning, which leverages unlabeled data through pretext tasks to improve data representation; Transfer Learning, which applies insights from one domain to improve predictions in another, effectively addressing the cold-start problem; Active Learning, which strategically queries labels for the most informative data points, optimizing the learning process in sparse data environments; and Online Learning, which continuously updates models in response to new data, ensuring adaptability and timeliness. Each technique uniquely addresses critical challenges such as data sparsity, scalability issues, and the integration of new users or items, thereby improving the overall effectiveness of recommendation systems.
	
	\subsection{Self-Supervised Learning}
	\label{sec:ssl}
	Self-supervised learning (SSL) has emerged as an essential strategy to address data scarcity in recent years. SSL represents a middle ground between supervised and unsupervised learning, wherein the algorithm generates its own supervised signals using pretext tasks and then learns useful representations beneficial for actual tasks. The pretext tasks commonly used in SSL include data completion, where the algorithm hides part of the data (a portion of a sentence or a region of an image) and then attempts to complete the missing part. Other pretext tasks involve denoising the data or reversing transformations on image data. SSL can be broadly divided into two classes: auto-associative SSL, which we have already discussed, involves the algorithm hiding part of the data and attempting to complete it or reconstruct the entire input. The second category is contrastive SSL, where the algorithm distinguishes between similar pairs of data points, referred to as positive, and dissimilar pairs, referred to as negative pairs. Positive pairs are generally obtained by applying a transformation or noise to a single data point, whereas negative pairs are obtained by randomly sampling from the data.
	
	SSL is particularly useful when unlabeled data are abundant. In supervised learning, such unlabeled data is often useless, but SSL makes good use of it by learning useful representations, which can then be fine-tuned on smaller labeled datasets. SSL can be beneficial in recommender systems \cite{yu2023selfsupervised, ren2024comprehensive}, where implicit feedback data are abundant in the form of clicks, views, and purchase history. This data can help build a user model that can be fine-tuned on labeled data, such as rating data.
	
	\subsubsection{Data Augmentation}
	Data augmentation is crucial in developing self-supervised learning methods for recommender systems. Data augmentation strategies can be divided into three main categories \cite{yu2023selfsupervised}, each tailored to address different aspects of SSL and its application in recommendation systems:
	\begin{itemize}
		\item \textbf{Sequence-Based Augmentation:} This involves modifying user interaction sequences to create varied learning examples \cite{s3rec,XieXu2022CLfS,ChengMingyue2021LTUR,hyperbolic-hypergraphs,LiuZhiwei2021CSSR}. Techniques include item masking, reordering, and substituting, which help models learn robust features by predicting the original order of items.
		
		\item \textbf{Graph-Based Augmentation:} Applied primarily in graph-based recommender systems, where a bipartite graph represents the interactions between users and items \cite{YuJunliang2022SMHC, JiancanWu2021SGLf,Junliang21, ZhouXin2023SASF,Yonghui21, YangHaoran2021HMCL}. These methods include edge dropping, node dropping, and edge adding. Such augmentations aim to simulate variations in user-item interaction graphs, enhancing the model's ability to capture essential connectivity patterns.
		
		\item \textbf{Feature-Based Augmentation:} Focuses on altering the feature representations of users or items, such as adding noise to features or modifying feature values  \cite{Ruihong22, QiuRuihong2021MAMC, KalantidisYannis2020HNMf, YuJunliang2022AGAN, LinZihan2022IGCF}. This approach is intended to make the recommendation models more adept at handling feature variability and improving their generalization capabilities.
	\end{itemize}
	Each of these augmentation strategies plays a crucial role in enriching the training data and enhancing the learning capacity of SSL models in recommender systems. 
	
	\subsubsection{Auo-associative Methods}
	Auto-associative SSL methods enable models to derive meaningful patterns from incomplete data, enhancing their predictive and generative capabilities. They consist of learning data representations by reconstructing entire inputs from corrupted ones or predicting missing parts of the data:
	\begin{itemize}
		\item \textbf{Structure generation:} Techniques like BERT4Rec \cite{BERT4Rec} and G-BERT \cite{shang2019pretraining} utilize masked item prediction and graph-based reconstructions, respectively. They involve masking parts of the input (such as items in sequences or nodes in graphs) and predicting these masked parts to learn robust data representations. 
		\item \textbf{Feature generation:} Approaches such as PMGT \cite{PMGT} and GPT-GNN \cite{GPT-GNN} focus on regenerating missing features or entire user/item profiles from partially available data, treating the task as a regression problem.
		\item \textbf{Sample prediction:} Techniques include enhancing sequence recommendation models by augmenting short sequences with pseudo-prior items or using semi-supervised learning methods to improve sample quality iteratively.
		\item \textbf{Pseudo-label prediction:} Involves using pre-defined relations or learned continuous values as labels for training. Models predict these relations or attempt to minimize the difference between the predicted and actual values, refining user-item interaction predictions.
	\end{itemize}
	
	Generative auto-associative methods leverage the power of architectures like Transformers for large-scale pre-training but face challenges related to computational demands. Auto-associative methods based on predictive pretext tasks offer dynamic and flexible sample generation but require careful design to ensure these tasks align with user-item interaction patterns.
	
	\subsubsection{Contrastive Methods}
	Contrastive pretext tasks derived from recommendation data augmentation approaches and data types cane be categorized into three groups \cite{yu2023selfsupervised}: structure-level contrast, feature-level contrast, and model-level contrast.
	\begin{itemize}
		\item \textbf{Structure-level contrast} utilizes user behavior data represented as graphs or sequences, exploiting slight perturbations to infer similar semantics. This is divided into same-scale and cross-scale contrasts:
		\begin{itemize}
			\item \textbf{Local-local contrast:} Focuses on graph-based models, maximizing mutual information between node representations from two augmented views using a shared encoder and contrastive loss (InfoNCE loss \cite{vandenOordAaron2019RLwC}). Key models include SGL \cite{JiancanWu2021SGLf}, DCL \cite{LiuZhuang2021CLfR}, and HHGR \cite{ZhangJunwei2022DSHL}, which employ various node and edge dropout techniques to enhance graph representation.
			\item \textbf{Global-global contrast:} Used in sequential recommendation models, where sequence augmentations are treated as global views and contrasted using a Transformer-based encoder. Notable implementations are CL4SRec \cite{XieXu2022CLfS}, H$^2 $SeqRec \cite{hyperbolic-hypergraphs}, and UniSRec \cite{tusr22}, utilizing methods like item masking and cropping.
			\item \textbf{Local-global contrast:} Aims to integrate global information into local structures, exemplified by EGLN\cite{Yonghui21} and BiGI \cite{bgev21}, which contrast user-item pair representations against global graph representations.
			\item \textbf{Local-context contrast:} Involves contrasting node or item sequences against their respective contextual clusters, with applications in models like NCL \cite{LinZihan2022IGCF} and ICL \cite{icl22} for capturing semantic neighbors or user intents.
		\end{itemize}
		
		\item \textbf{Feature-level contrast} leverages a variety of categorical features. SL4Rec \cite{Tiansheng21} and SLMRec \cite{TaoZhulin2023SLfM} are notable for applying correlated feature masking and dropout to augment data meaningfully.
		
		\item \textbf{Model-level contrast} modifies the model architecture itself to create augmented views dynamically. Techniques include neuron masking and adjusting hidden representations, with DuoRec \cite{Ruihong22} and SimGCL \cite{YuJunliang2022AGAN} as key examples that enhance model robustness and mitigate popularity bias.
	\end{itemize}
	
	The contrastive loss, essential for optimizing the mutual information between representations, includes popular estimators like Jensen-Shannon and InfoNCE \cite{vandenOordAaron2019RLwC}. These losses are crucial for learning distinct representations and managing negative sampling in contrastive learning frameworks.
	
	Despite the rapid expansion of contrastive methods in recommender systems, challenges remain such as the lack of rigorous understanding of augmentation impacts and the potential negative effects of common augmentations on model performance \cite{yu2023selfsupervised}.
	
	Self-supervised learning (SSL) significantly enhances recommender systems by enabling the learning of representations without labeled data. Innovative data augmentation techniques and advanced model architectures play a crucial role in improving the accuracy and efficiency of these systems. Despite being a relatively new field that has attracted considerable interest in recent years, SSL in recommender systems holds substantial potential and presents numerous unresolved challenges \cite{yu2023selfsupervised, ren2024comprehensive}.
	
	\subsection{Active Learning}
	\label{sec:active-learning}
	Active learning is a type of machine learning that selectively queries the data source to label new data points to achieve greater accuracy using strategies such as uncertainty sampling, estimated error reduction, or density-weighted methods \cite{settles2009active}. This approach is practical when labeled data is scarce or expensive, making it particularly useful for recommender systems settings.
	
	Guan et al. \cite{guan2017matrix} proposed the Enhanced Singular Value Decomposition (ESVD) model, which integrates the basic matrix factorization technique of Regularized Singular Value Decomposition (RSVD) \cite{koren2008factorization} with a rating completion strategy inspired by active learning. The strategy involves selecting the top $N$ most popular items and active users to obtain the densest sub-matrix, effectively reducing the sparsity problem. This densest sub-matrix is then used with the RSVD model to predict ratings, and the missing ratings in the original matrix are filled in with the ratings obtained from the sub-matrix. Finally, RSVD is applied again to the original matrix for a comprehensive rating prediction. Additionally, they extended the ESVD model to the Multilayer ESVD (MESVD), learning the model iteratively to achieve better performance. The output generated by the lower layer was used as input for the upper layer to obtain a much denser sub-matrix. All estimated ratings were filled in the original matrix to evaluate the model's performance.
	
	The authors further proposed two extensions of the ESVD to handle imbalanced datasets. The first was Item-wise ESVD (IESVD), which selects the top $N$ most popular items to form a sub-matrix and then chooses only the active users. The second extension was User-wise ESVD (UESVD), which selects the top $N$ most active users to form a sub-matrix and then chooses the most popular items from this sub-matrix.
	
	Experiments were conducted using several datasets: MovieLens-100k, MovieLens-1M, and Netflix, employing the Root Mean Square Error (RMSE) metric \cite{herlocker2004evaluating}. They compared the ESVD with MESVD, and the MESVD demonstrated a minor improvement in prediction accuracy with four layers.
	
	\subsection{Online Learning}
	\label{sec:online-learning}
	Online learning is a machine learning paradigm where data is sequentially accessible, allowing systems to update predictions progressively as new information is acquired \cite{ShalevShwartz2012OnlineLA, orabona2023modern}. This method is necessary when training over a complete dataset is computationally prohibitive or when a system needs to be deployed immediately and learn from data generated in real-time. Incremental learning ensures the system remains functional and progressively improves, providing timely and increasingly accurate responses without initial exhaustive data.
	
	A prominent application of online learning in recommender systems is in news recommendation. News recommender systems are distinct from other recommender systems due to unique challenges such as scalability, the transient nature of news, and dynamically changing user preferences \cite{li-2011}. These systems must effectively manage an enormous volume of news articles available online. Unlike static content like movies, news articles have a short relevance lifespan and need to be updated or replaced frequently. Therefore, these systems are designed to handle new and trending articles efficiently while adapting to changes in user interest over time.
	
	News article recommendations can be modeled using a contextual multi-armed bandit (MAB) framework. Each article is considered an arm in this model, and the reward is quantified by how frequently users click the article. This approach helps balance the exploration of new articles against the exploitation of previously popular articles, aiming to maximize user engagement through an optimal recommendation strategy. The core challenge in MAB problems, the exploration-exploitation trade-off, necessitates algorithms that can continuously assess the potential of new articles against the known rewards of familiar ones \cite{levinthal-1993}.
	The MAB problem involves an agent making sequential choices from a set of options, each with its reward distribution, to maximize cumulative rewards. For news recommendations, this translates into monitoring and adapting to user preferences in real-time, which makes MAB approaches particularly suitable for the dynamic and voluminous nature of news consumption.
	
	The basic MAB problem is formulated as follows. Each arm (article) in a set $A_t \in \{1, \dots, K\}$ is associated with unknown reward distributions $\{D_1, \dots, D_K\}$ and mean rewards $\{\mu_1, \dots, \mu_k\}$. The agent selects an arm $a(t)$ at each time step $t = \{1, 2, \dots\}$, and observes a reward $r_{a(t)}$. The primary goal is to minimize the total regret $R_T$ over a predetermined number of trials $T$, where regret is defined as the difference between the total reward obtained and the maximum possible reward:
	
	\begin{equation}
		R_T = T \mu^* - \sum_{t=1}^T \mu_{a(t)},
	\end{equation}
	where $\mu^* = \max_{i=1, \dots, k} \mu_i$ is the highest expected reward achievable by any arm, and $\mu_{a(t)}$ is the reward received from arm $a(t)$ at time $t$. This framework emphasizes the necessity to balance exploring new articles and exploiting known successful ones to effectively cater to user preferences and maximize engagement \cite{gupta-2006}.
	
	In the realm of Multi-Armed Bandits (MAB), the primary challenge is to effectively manage the \textbf{exploration-exploitation dilemma}, where the goal is to maximize rewards by exploiting known rewards or exploring new potentially rewarding actions. The strategies are categorized into context-free and contextual algorithms, each suited for different scenarios regarding available information about the environment and the arms.
	
	\begin{itemize}
		\item \textbf{Context-free bandit algorithms}: make decisions based solely on historical rewards of each action, without considering any specific attributes of the arms. Some examples of this type of algorithms are:
		\begin{itemize}
			\item \textbf{Epsilon-greedy Algorithm:} This approach selects the best-performing arm with a probability of $1 - \epsilon$ and any other arm with a probability of $\epsilon$. It introduces a straightforward method to balance exploiting the best arm and exploring others, though managing the value of $\epsilon$ is critical for its effectiveness \cite{vermorel-2005, cesa-1998}.
			
			\item \textbf{Boltzmann Exploration (SoftMax):} It selects arms based on their expected rewards adjusted by a temperature parameter $\tau$, which regulates the exploration level. The selection probabilities are derived from the Gibbs or Boltzmann distribution, making it more refined than epsilon-greedy by focusing on promising arms more frequently \cite{cesa-1998}.
			
			\item \textbf{Upper Confidence Bound (UCB):} This method selects arms based on their average rewards and the uncertainty or variance in their rewards. The critical aspect of UCB is its use of an upper confidence bound to balance exploitation and exploration naturally, favoring arms that are potentially under-explored \cite{auer-2002}.
		\end{itemize}
		
		\item \textbf{Contextual bandit algorithms}: use additional context or attributes to improve decision accuracy and effectiveness. This can include user profiles, article descriptions, or any relevant features that might influence decision-making. Some contextual algorithms are:
		\begin{itemize}
			\item \textbf{Epoch-Greedy:} Operates by selecting arms using a set of hypotheses about the rewards, which could be based on user features or past interactions. It alternates between exploration by randomly choosing arms and exploitation by selecting the best arm according to the current hypothesis \cite{kuleshov-2014, langford-2008}.
			
			\item \textbf{LinUCB:} A more sophisticated approach that uses linear regression to estimate the rewards associated with each arm's features. It updates its estimates of the arm's rewards based on observed payoffs and adjusts the exploration-exploitation balance using a confidence bound on the estimated rewards \cite{li-2010}.
		\end{itemize}
	\end{itemize}
	
	Following extensive experimentation on the Yahoo! Front Page Today Module dataset, which contains over 33 million events, the authors in \cite{li-2010} conclude that contextual algorithms outperform non-contextual ones in scenarios where user and content dynamics are constantly changing, such as web services. Specifically, the new contextual bandit algorithm showed a 12.5\% improvement in click lift over a standard context-free bandit algorithm. This performance enhancement was even more pronounced when the data were sparser. By using additional information about the environment and entities involved, contextual algorithms can adapt more effectively to these dynamics. They provide a more personalized experience and can achieve higher performance metrics, such as click-through rates, by tailoring decisions based on the context of each situation.
	
	\subsection{Transfer Learning}
	\label{sec:transfer-learning}
	Transfer learning is a powerful machine learning technique that involves reusing a pre-trained model as a starting point to develop another model for a new task \cite{Yang_Zhang_Dai_Pan_2020}. This approach can help improve learning efficiency and accuracy in a new task by leveraging knowledge and data from a related task that has already been mastered. Transfer learning is particularly useful when there is a shortage of data available for the new task. It is widely used in several fields, including computer vision, natural language processing, and recommender systems \cite{zhuang2020comprehensive}.
	
	Traditional recommender systems, such as factorization-based collaborative filtering, require extensive training datasets to work effectively but encounter difficulties when dealing with sparse real-world data and the cold-start problem related to new users or items. Transfer learning methods can alleviate this limitation. These methods include instance-based and feature-based approaches that aim to enhance recommendations. 
	\begin{itemize}
		\item \textbf{Instance-based methods} transfer different data types, such as ratings or feedback, from one domain to another to improve recommendations. For example, Pan et al. \cite{PanWeike2021TLiC} use uncertain ratings from a source domain as constraints to aid in completing rating matrix factorizations in a target domain. Similarly, Hu et al. \cite{hu2019} employ an attentive memory network to extract and transfer helpful information from unstructured texts.
		
		\item \textbf{Feature-based methods}, on the other hand, transfer latent feature information across domains. Pan et al.'s Coordinate System Transfer (CST) \cite{PanWeike2010TLiC} uses user and item features from a source domain. It applies them as constraints in a target domain to improve recommendation accuracy significantly compared to non-transfer methods. 
		
		\item \textbf{Model-based methods} involve extracting common knowledge from a source domain and transferring it to a target domain. The goal is to transfer high-level rating behaviors, such as user and item clusters or memberships, which can help alleviate the sparsity problem in the target domain. Several algorithms have been proposed, including CBT \cite{li-2009a}, RMGM \cite{li-2009b}, CLFM \cite{GaoSheng2013CRvC}, CKT-FM \cite{PanWeike2015Cktv}, and DSNs \cite{KanagawaHeishiroCRvD}. 
	\end{itemize}
	
	Further studies explore cross-domain recommendations using advanced techniques like Bayesian neural networks and deep learning frameworks for feature mapping and domain adaptation, further enhancing the effectiveness of recommender systems across varied data sparsity levels \cite{PanWeike2013Tlih, Zhen2020SRvC, ZhuangFuzhen2018Tcff}. Most models are primarily designed to enhance their predictive capabilities by incorporating knowledge from direct user interactions, such as quiz responses, ratings with varying levels of certainty, and straightforward like/dislike feedback.
	
	\subsection{Discussion}
	This section has reviewed essential learning strategies in recommendation systems, addressing critical challenges like data sparsity and the cold-start problem. Self-Supervised Learning (SSL) utilizes unlabeled data for learning through pretext tasks, enhancing model capability without explicit feedback. Transfer Learning applies knowledge from one domain to improve performance in another, proving vital for domain-specific challenges. Active learning focuses on selectively querying labels for the most informative data points and optimizing resource use in sparse data scenarios. Online learning updates models incrementally with new data, which is essential for adapting to real-time changes in user preferences. Together, these strategies improve the effectiveness and efficiency of recommendation systems, each offering unique solutions to the complexities of modeling user-item interactions.
	
	Unlike domains such as natural language processing (NLP) and computer vision, where massive, freely accessible datasets are abundant, recommendation systems often grapple with a scarcity of openly available data. The proprietary nature of recommendation data presents significant hurdles. Companies guard their user interaction data as a valuable asset, often hesitating to share it due to competitive and privacy concerns. Moreover, recommendation data is frequently highly specialized, targeting specific applications like hotel bookings, online shopping, or media streaming, which amplifies the difficulty of generalizing findings across different domains. These challenges underscore the critical importance of transfer learning and domain adaptation in recommendation systems. Transfer learning allows the application of models developed in one domain to be adapted for use in another, leveraging learned patterns and knowledge even when direct data transfer is impossible. Although a relatively new research area within recommendation systems, transfer learning has gained considerable attention recently and is poised for continued growth and development. Its potential to overcome data scarcity and specialization challenges in recommendation systems is significant, making it a promising area for future research and development.
	
	Another significant challenge researchers face in recommender systems is the reluctance of users to provide explicit ratings, which are crucial for training traditional supervised learning models. However, an abundance of implicit data, such as clicks, views, or purchase histories, can be harnessed to enhance model training. SSL, a concept that has sparked significant interest across various domains, including recommendation systems, offers a promising solution. It uses unlabeled data to generate labels through pretext tasks, opening up substantial potential for SSL in recommendation systems to effectively utilize abundant implicit data. Despite the remaining challenges, including the need for a more rigorous theoretical foundation and the development of sophisticated,  domain-specific data augmentation methods, the future of the field appears promising.
	
	Federated learning is an important learning strategy not covered in detail in this section. It is an approach to decentralized machine learning in which multiple devices work together to train a shared model while keeping the data localized on each device \cite{YangQiang2019FMLC}. This method enhances data privacy and security, as the raw data never leaves the local devices. In the context of recommender systems, federated learning is essential because it allows creating personalized recommendations without compromising user privacy \cite{SunZehua2024ASoF}. It also enables scalability by leveraging the computational power of numerous devices and reduces latency by minimizing data transfers.
	
	As the field progresses, the focus on refining SSL approaches for better utilization of implicit data and broadening the scope of transfer learning to mitigate the challenges of data scarcity and specialization in recommendation systems will undoubtedly continue to be key research areas. Furthermore, as privacy concerns and data security become increasingly critical, federated learning also constitutes a promising research direction in developing recommendation technologies. These strategies not only promise to enhance the accuracy and effectiveness of recommendation models but also strive to make them more adaptable and robust across varying domains and scarce data environments.
	
	\section{Optimization}
	\label{sec:optimization}
	Optimization is vital in the development and effectiveness of latent factor models used in recommender systems. The quality of the resulting model depends heavily on the optimization algorithm used, as recommendation data is usually voluminous and complex. This complexity requires significant computational resources in terms of processing time and memory. Efficient optimization algorithms are, therefore, crucial as they allow the development, testing, and deployment of more powerful models. These advanced models can capture complex patterns within the data, ultimately improving the quality of recommendations provided to users.
	
	\cite{RuderSebastian2017Aoog, SunShiliang2020ASoO, SunRuo20}
	
	\subsection{Stochastic Gradient Descent}
	\label{sec:sgd}
	Gradient descent requires computing the gradient using the entire training set, which can be computationally expensive for large datasets. However, since the loss function is the summation of individual example losses, we can approximate the total loss using a subset $\mathcal{B}$ of the training set containing $b$ examples:
	\begin{equation}
		\mathcal{B}=\{r_{u_1 i_1}, \ldots, r_{u_b i_b}\}  \subset \mathcal{R}.
	\end{equation}
	The subset $\mathcal{B}$ is called a \emph{mini-batch}, and its size, $b$, is a critical hyperparameter of the algorithm. The loss associated with the batch $\mathcal{B}$ is:
	\begin{equation}
		L^{\mathcal{B}}(\theta)= \frac{1}{b}\sum_{r_{ui} \in \mathcal{B}} \ell(\hat{r}_{ui}, r_{ui}),
	\end{equation}  
	and serves as a proxy for the total loss $L$, resulting in the objective function:
	\begin{equation}
		J^{\mathcal{B}}(\theta) = L^{\mathcal{B}}(\theta) + \lambda \Omega(\theta).
	\end{equation} 
	The update rule for stochastic gradient descent can then be written as:
	\begin{equation}
		\theta = \theta - \eta \nabla_{\theta} J^{\mathcal{B}}(\theta).
	\end{equation}
	The algorithm thus obtained is called Stochastic Gradient Descent (SGD) \footnote{Some authors refer to the gradient descent algorithm that uses the total gradient as batch gradient descent and use the name Mini-batch Gradient Descent when the update is done using a mini-batch. They reserve the name SGD to the particular case when the batch size is 1.}. Although this change compared to gradient descent seems trivial, it has profound implications, both theoretical and practical, on the behavior of the algorithm and its effects on real applications:
	\begin{itemize}
		\item A single complete pass over the data, by sampling enough mini-batches, is called an \emph{epoch}. Typically, several epochs are necessary for the algorithm to converge. SGD reduces the cost of each step, which means it performs several updates in a single epoch, whereas gradient descent updates the model parameters only once per epoch. For many problems, this leads to accelerated convergence \cite{Nemirovski09, Agarwal12}, and for large datasets, the algorithm might reach an acceptable solution before passing through the entire dataset.
		
		\item Unlike the total gradient, where the objective value decreases at each step (for an appropriate choice of the learning rate), using a mini-batch results in fluctuations of the objective. This property can be beneficial in some cases as it allows the algorithm to skip shallow local minima in a way similar to simulated annealing. The noisy gradient, however, can lead to slow convergence even when the algorithm is exploring a good local minimum. In the extreme case of a single example, a small mini-batch size results in fast but very noisy updates, whereas large mini-batches reduce noise as the partial gradient aligns more with the total gradient. Large batch sizes, however,  result in a higher computational cost and may lead the algorithm into shallow local minima near the initial position. The size of the mini-batch is, therefore, not only a computational parameter that controls the time and memory required to find a solution, but it is also a learning parameter that affects the quality of the solution found.
		
		\item The gradient of a mini-batch does not always vanish at a minimum, unlike the total gradient. The mini-batch estimates have a large variance,  meaning that each update might point in a slightly different direction. As a result, the parameters can oscillate around the minimum instead of smoothly converging. This phenomenon could prevent the algorithm from reaching the minimum because the parameters keep oscillating indefinitely. Therefore, it is necessary to gradually decrease the learning rate, which can be achieved using a learning schedule \cite{Darken92} or by decay, such as linear or exponential decay.
	\end{itemize}
	The Stochastic Gradient Descent (SGD) algorithm and its variants, including momentum-based and adaptive learning rate algorithms, have been pivotal in advancing deep learning. These optimization techniques are essential for effectively training large-scale deep models on massive datasets, enabling the practical implementation and success of complex neural network architectures.
	
	\subsection{Momentum Methods}
	\label{sec:momentum}
	SGD updates the model parameters using the gradient computed on a randomly sampled mini-batch, which is computationally efficient but introduces some challenges. The gradients of mini-batches can point in various directions different from the direction of the total gradient, leading to noisy updates that slow down convergence. Additionally, the objective function in learning tasks often has a landscape of deep valleys with flat basins. In the steep regions of the valley, gradient updates result in a rapid reduction of the objective function. However, SGD struggles to progress in the flat basins as the gradient diminishes.
	
	The momentum method \cite{POLYAK19641} addresses these issues by using a smoothed version of the gradient to update the model parameters, thereby reducing noisy updates. The algorithm maintains an exponentially decaying average of the previous gradients through a variable named $ v $ (for velocity) and uses it as the descent direction:
	\begin{align}
		v &= -\eta \nabla_{\theta} J^{\mathcal{B}}(\theta) + \alpha v, \\
		\theta &= \theta + v,
	\end{align}
	where $0 \leq \alpha \leq 1$ is the momentum coefficient. The value of $\alpha$ determines how much past gradients influence the current update. A higher $\alpha$ (close to 1) gives more weight to past gradients, effectively smoothing the parameters' trajectory and helping dampen oscillations. This can be particularly beneficial in the flat regions of the objective function, as it allows the algorithm to maintain momentum and make steady progress. Conversely, a lower $\alpha$ (close to 0) relies more on the current gradient, making the updates more responsive to the immediate gradient but potentially increasing the noise and oscillations.
	
	Nesterov's momentum method \cite{nestrov}, also known as Nesterov's Accelerated Gradient (NAG), is an improvement over the original momentum method where the gradient is computed using the updated parameters instead of the current ones:
	\begin{align}
		\tilde{\theta} &= \theta + \alpha v,\\
		v &= -\eta \nabla_{\theta} J^{\mathcal{B}}(\tilde{\theta}) + \alpha v, \\
		\theta &= \theta + v,
	\end{align}
	The lookahead included in the computation of the descent direction provides additional information that helps anticipate the future position of the parameters, which reduces oscillations and speeds up convergence.
	
	\subsection{Adaptive Learning Rate Methods}
	\label{sec:alr}
	The learning rate is a crucial parameter for all gradient descent-type algorithms, particularly stochastic gradient descent ones. Inappropriate choices of the learning rate can cause slow convergence, oscillations, or even divergence. Using predefined learning schedules \cite{Darken92} or simple decay strategies can mitigate this issue, but these require extensive tuning and domain-specific knowledge.
	
	Adaptive learning rate methods provide efficient tools to overcome these limitations by automatically adjusting the learning rate. This adaptation operates along two dimensions: individually for each model parameter and over the course of the optimization process. Adapting to model parameters stems from the understanding that not all parameters have the same impact on the objective function. Certain parameters may have a greater influence on the objective function due to the significant effect of their corresponding features on the model's output. Consequently, small changes in such parameters can lead to significant shifts in the objective value. Using the same learning rate for all parameters can thus be problematic. On the one hand, a small learning rate is necessary to prevent overshooting sensitive parameters, but it can lead to slow convergence for less sensitive ones. On the other hand, a large learning rate can cause the opposite effect.
	
	The Adaptive Gradient (AdaGrad) algorithm \cite{adagrad} remedies this issue by using a different learning rate for each model parameter. It adjusts these learning rates individually by scaling them inversely proportional to the square root of the sum of all historical squared values of the gradient of their respective parameters.
	\begin{align}
		g &= \nabla_{\theta} J^{\mathcal{B}}(\theta),\\
		r &= r + g \odot g,\\
		v &= -\eta\frac{g}{\delta + \sqrt{r}}, \quad \text{(element-wise)} \\
		\theta &= \theta + v,
	\end{align}
	where $r$ is a vector that stores the sum of squared partial derivatives of all model parameters, $\odot$ denotes element-wise multiplication, and $\delta$ is a small number used for numerical stability.
	
	As a result of this strategy, parameters with the largest partial derivative of the loss experience a rapid decrease in their learning rate, while parameters with small partial derivatives have a relatively small decrease. This ensures greater progress in the more gently sloped directions of the parameter space.
	
	Root Mean Square Propagation (RMSProp) \cite{rmsprop} is an improvement over AdaGrad that adjusts the learning rate during the optimization process. While AdaGrad works well for convex functions, it can be less effective for non-convex functions with a complex landscape. In such cases, the sensitivity of the objective function to the model parameters may drastically change from one region to another. Consequently, old values of $ r $ can be misleading and hinder the algorithm's progress. RMSProp addresses this issue by giving more weight to recent gradients through the use of an exponentially weighted moving average instead of a cumulative sum:
	\begin{align}
		g &= \nabla_{\theta} J^{\mathcal{B}}(\theta),\\
		r &= \rho r + (1-\rho) g \odot g,\\
		v &= -\eta\frac{g}{\delta + \sqrt{r}}, \quad \text{(element-wise)} \\
		\theta &= \theta + v,
	\end{align}
	where $ 0 \leq \rho \leq 1 $ is the decay rate or smoothing factor.
	
	Using an exponentially decaying average helps maintain efficient learning rates throughout the optimization process. Typically, $\rho$ is chosen empirically. A smaller $\rho$ gives more weight to recent gradients and can be beneficial for rapidly changing objectives. In contrast, a larger $\rho$ provides a smoother update and can be useful for more stable objectives.
	
	Adaptive Moment (Adam) \cite{adam} is another adaptive learning rate algorithm that combines RMSProp and the momentum algorithm. It uses two momentum vectors: one for the gradient, denoted by $ r $, and another for the squared partial derivatives, denoted by $ s $. Before updating the model parameters, Adam performs bias-correction terms for these moments, ensuring more precise estimates, especially during the initial stages of training.
	
	\begin{align}
		g &= \nabla_{\theta} J^{\mathcal{B}}(\theta),\\
		s &= \rho_1 s + (1-\rho_1) g,\\
		r &= \rho_2 r + (1-\rho_2) g \odot g,\\
		\tilde{s} &= s / (1-\rho_1^t),\\
		\tilde{r} &= r / (1-\rho_2^t),\\
		v &= -\eta \frac{\tilde{s}}{\delta + \sqrt{\tilde{r}}}, \quad \text{(element-wise)} \\
		\theta &= \theta + v,
	\end{align}
	where $ \rho_1, \rho_2  \in (0,1)$ are smoothing factors, and $ t $ is the time step (starting from 1). Note that as the number of steps increases, $ \rho_1^t$ and $\rho_2^t$ tend to zero, and the bias correction practically vanishes. The blend of the strategies of the momentum algorithm and RMSProp allows Adam to efficiently adapt to variations in the behavior of the objective function. This adaptability allows Adam to navigate the complex landscapes of the objective functions typically encountered with large, complex learning models, making it one of most widely used algorithms for training large models, whether in latent factor models or deep neural models.  
	
	Adaptive Moment Estimation (Adam) \cite{adam} is another adaptive learning rate algorithm that combines RMSProp and the momentum algorithm. It uses two momentum vectors: one for the gradient, denoted by $ s $, and another for the squared partial derivatives, denoted by $ r $. Before updating the model parameters, Adam performs bias correction for these moments, ensuring more precise estimates, especially during the initial stages of training.
	
	\begin{align}
		g &= \nabla_{\theta} J^{\mathcal{B}}(\theta),\\
		s &= \rho_1 s + (1-\rho_1) g,\\
		r &= \rho_2 r + (1-\rho_2) g \odot g,\\
		\tilde{s} &= s / (1-\rho_1^t),\\
		\tilde{r} &= r / (1-\rho_2^t),\\
		v &= -\eta \frac{\tilde{s}}{\delta + \sqrt{\tilde{r}}}, \quad \text{(element-wise)} \\
		\theta &= \theta + v,
	\end{align}
	where $ \rho_1, \rho_2 \in (0,1) $ are smoothing factors, and $ t $ is the time step. Note that as the number of steps increases, $ \rho_1^t $ and $ \rho_2^t $ tend to zero, and the bias correction practically vanishes. The combination of the momentum algorithm and RMSProp allows Adam to efficiently adapt to significant variations in the behavior of the objective function. This adaptability enables Adam to navigate the complex landscapes of the objective functions typically encountered in large, complex learning models, making it one of the most widely used algorithms for training large models, whether in latent factor models or deep neural networks.
	
	The momentum method and Adam can be seen as gradient variance reduction techniques whereby the gradient is smoothed out to maintain a consistent descent direction. Other methods that have been proposed in this direction are Stochastic Average Gradient (SAG) \cite{sag} and Stochastic Variance Reduced Gradient (SVRG) \cite{svrg}, which both aim to reduce the variance of the gradient estimates to achieve faster and more stable convergence.
	
	\subsection{Dedicated Methods}
	\label{sec:dedicated}
	In contrast to the general optimization algorithms discussed in the previous sections, which are widely used across various machine learning problems, including neural network training, this section focuses on specialized algorithms explicitly tailored for latent factor models in recommender systems. By leveraging the specific structure and characteristics of recommender system data, these methods enhance latent factor models' computational efficiency and predictive performance.
	
	Among these specialized methods, Alternating Least Squares (ALS) \cite{als} is an optimization technique widely used for matrix factorization. The term "alternating" refers to the optimization process, which alternates between fixing one set of variables while optimizing another set. In the case of matrix factorization, this typically involves holding user factors constant to solve for item factors and vice versa. ALS uses the least squares approach to minimize the squared differences between observed and predicted ratings, adjusting factors to fit the data as closely as possible. To solve the optimization problem associated with the matrix factorization model:
	\begin{equation} 
		\min_{\{p_u\}, \{q_i\}} \dfrac{1}{2} \sum_{r_{ui} \in \mathcal{R}}\left (p_u^T q_i - r_{ui}\right )^2 + \dfrac{\lambda}{2} \left(\sum_{u=1}^n \|p_u\|_2^2+ \sum_{i=1}^m \|q_i\|_2^2\right),
	\end{equation}
	ALS proceeds as follows:
	\begin{enumerate}
		\item Initialize $\{q_i\}$ randomly.
		
		\item With $\{q_i\}$ fixed, solve for $\{p_u\}$ the following optimization problem:
		\begin{equation}
			\min_{\{p_u\}} \dfrac{1}{2} \sum_{r_{ui} \in \mathcal{R}}\left (p_u^T q_i - r_{ui}\right )^2 + \dfrac{\lambda}{2} \sum_{u=1}^n \|p_u\|_2^2.
		\end{equation}
		This linear least squares problem can be solved using various efficient methods \cite{LawsonCharlesL1974Slsp}, including direct linear algebraic methods.
		
		\item With $\{p_u\}$ found in the previous step fixed, solve for $\{q_i\}$ the following optimization problem:
		\begin{equation}
			\min_{\{q_i\}} \dfrac{1}{2} \sum_{r_{ui} \in \mathcal{R}}\left (p_u^T q_i - r_{ui}\right )^2 + \dfrac{\lambda}{2} \sum_{i=1}^m \|q_i\|_2^2.
		\end{equation}
		This is also a linear least squares problem that can be solved similarly to the previous problem.
		
		\item Repeat Steps 2 and 3 until convergence.
	\end{enumerate}
	
	Each subproblem that results from fixing either user or item factors is a simple linear least squares problem that can be solved efficiently using several efficient algorithms and highly optimized linear algebra libraries. The ability to solve these subproblems efficiently renders the overall algorithm highly efficient.
	
	To enhance the computational efficiency of SVD++ \cite{koren2009matrix}, Wang et al. \cite{wang2020svd++} introduced an improved Singular Value Decomposition++ (SVD++) that incorporates a Backtracking Line Search \cite{royer2018complexity} in the SVD++ algorithm (BLS-SVD++). This approach accelerates SVD++ and improves prediction accuracy by employing a backtracking line search strategy to determine the optimal step size along a particular descent direction. This optimization is based on the local gradient of the objective function. Their experiments utilized the MovieLens-1M, MovieLens-10M, and FilmTrust datasets. The BLS-SVD++ model was compared with traditional SVD \cite{koren2009matrix}, regularized SVD (RSVD) \cite{koren2008factorization}, and SVD++ \cite{koren2009matrix}. The results demonstrate the effectiveness of integrating a backtracking line search within the SVD++ algorithm, significantly reducing the number of iterations and enhancing prediction accuracy.
	
	Nasiri \cite{nasiri2016increasing} addressed the challenges of convergence speed and data sparsity by proposing a novel method for the optimization-based matrix factorization technique, which serves as a preprocessing step to initialize the latent factor matrices of users and items. The proposed method consists of two parts: First, they employed the Singular Value Decomposition (SVD) method to decompose the user-item matrix into component matrices. These matrices were then used as initial values for the latent factor matrices in the Stochastic Gradient Descent (SGD) technique \cite{funk2006netflix}, facilitating faster algorithm convergence. The authors conducted experiments on the MovieLens-100k dataset and used RMSE to evaluate the model's performance. They compared the performance of SGD with initialization against SGD without initialization. The results showed that initializing SGD significantly improved prediction accuracy and reduced the number of iterations required to reach a minimal error rate.
	
	\subsection{Discussion}
	Efficient optimization algorithms play a crucial role in developing effective latent factor models that provide high-quality recommendations to users. Stochastic Gradient Descent (SGD) is a key algorithm in this context, enabling the training of large-scale models on massive datasets. Using mini-batches in SGD allows for accelerated convergence and the exploration of good local minima, but it also introduces challenges, such as noisy updates and oscillations around the minimum. To address these issues, several variants of SGD have been proposed, including momentum-based and adaptive learning rate algorithms such as AdaGrad, RMSProp, and Adam. These optimization techniques are fundamental for practically implementing large latent factor models, particularly those relying on deep neural networks. In addition to these general-purpose algorithms, we present algorithms specifically designed for collaborative filtering, considering the unique structure of the recommender system's latent factor models to solve the associated optimization problem efficiently.
	
	Optimization for latent factor models and, more generally, for machine learning is a vast field, and this section does not cover all existing algorithms. An important class of algorithms not covered in our presentation is high-order algorithms \cite{Nocedal2018NumericalO}, such as Conjugate Gradient \cite{Hestenes1952MethodsOC,hager2005new}, Quasi-Newton methods \cite{BROYDEN1970TCoa, FletcherR}, and Limited-memory Broyden-Fletcher-Goldfarb-Shanno algorithm (L-BFGS) \cite{NocedalJorge1980Uqmw, LIUDC1989Otlm}, which also play significant roles in various machine learning and recommender systems applications. High-order methods use curvature information obtained from second derivatives to accelerate convergence. They require fewer steps than first-order methods, but each step involves more complex computations and takes more time.
	
	The main issue faced in optimization today is that, unlike earlier models, such as Matrix Factorization, which were convex, recent models are far more complex and often lead to non-convex optimization problems. This is a general trend in machine learning, particularly emphasized by the widespread adoption of deep learning models. Non-convex optimization has long been recognized as a challenging problem compared to convex optimization. Researchers have known for decades that the main line separating easy from difficult optimization problems is not linear versus nonlinear but rather convex versus non-convex problems. However, there is still insufficient work in non-convex optimization and no efficient algorithms for large-scale non-convex problems. The rise of deep learning and the interest it has attracted from industry and governments can be a significant factor in driving deeper exploration into the difficult yet promising field of non-convex optimization.
	
	Another important research direction involves analyzing the behavior of the various optimization algorithms with latent factor models in collaborative filtering. Recommendation data is typically sparse, high-dimensional, and often noisy, which presents unique challenges for optimization algorithms. An in-depth analysis of how these algorithms perform under such conditions can help researchers gain valuable insights into their effectiveness and limitations.
	By understanding the strengths and weaknesses of these algorithms in the context of recommender systems, researchers and practitioners can make more informed decisions, potentially improving the performance of recommendation engines.
	Moreover, these insights can also help design more specialized algorithms specifically crafted for latent factor models in collaborative filtering. Although some specialized algorithms exist, further effort is needed to enhance the efficiency and accuracy of recommendations.
	
	\section{Conclusion} 
	This survey systematically reviews the latest techniques and advancements in latent factor models for recommender systems. It covers various aspects such as learning data, model architecture, learning strategies, and optimization techniques, providing a thorough understanding of the field. The survey not only highlights strengths, identifies trends, and points out gaps in current research, but also demonstrates the effectiveness of latent factor models in addressing challenges like data sparsity and scalability in recommendation tasks.
	
	Through this analysis, we provided insights into how different machine-learning paradigms can enhance recommender systems. We discussed potential future research directions, emphasizing the need for more robust, adaptable, and context-aware recommendation methods.
	
	By surveying and categorizing existing methodologies, this survey aims to guide researchers and practitioners in developing more effective and personalized recommender systems.

	\section*{Acknowledgments}
	This research work is supported by the Research Center, CCIS, King Saud University, Riyadh, Saudi Arabia.
	
	\bibliographystyle{IEEEtran}
	
\end{document}